\documentclass{svproc}
\usepackage{graphicx}
\usepackage{multirow}%
\usepackage{amsmath,amssymb,amsfonts}%
\usepackage{mathrsfs}%
\usepackage[title]{appendix}%
\usepackage{xcolor}%
\usepackage{textcomp}%
\usepackage{manyfoot}%
\usepackage{cleveref}
\usepackage{booktabs}%
\usepackage{subfig}
\usepackage{algorithm}
\usepackage{algorithmicx}%
\usepackage{algpseudocode}%
\usepackage{listings}%
\usepackage{bm}
\usepackage{url}

\begin{document}
\mainmatter              
\title{A Network Science Perspective on Evaluating Deep Graph Generative Models}
\titlerunning{Evaluating Deep Graph Generative Models} 
\author{Tianrui Mao\inst{1} \and Abele M\u{a}lan\inst{2} \and
Megha Khosla\inst{1} \and Lydia Chen\inst{1,2} \and Huijuan Wang\inst{1}}
\authorrunning{Tianrui Mao et al.} 
\institute{Delft University of Technology, the Netherlands
\and
University of Neuch\^{a}tel, Switzerland}

\maketitle                 

\begin{abstract}
Traditional network models from network science, such as the Erd\H{o}s--R\'{e}nyi and configuration models, generate random networks that reproduce a few selected topological properties observed in real-world networks. Deep graph generative models emerge as a data-driven approach, leveraging deep neural network architectures to learn complex structural distributions directly from real-world networks to generate more realistic synthetic networks. Because real social contact networks cannot be shared due to privacy risks, synthetic networks serve as an alternative for developing and evaluating epidemic mitigation strategies. In this work, we evaluate deep graph generative models from a network science perspective by assessing both the topological similarity between generated and real-world networks and their utility in identifying effective node immunization strategies to suppress epidemic/misinformation spreading.
It is found that two deep graph generative models produce synthetic networks that closely resemble the structural properties of real-world networks, enabling them to identify effective immunization strategies.

\keywords{Deep graph generative models, network models, Epidemic immunization, Social networks, Model evaluation.}
\end{abstract}
\section{Introduction}\label{sec1}
%

In network science, network models have been proposed to generate random graphs that preserve selected topological properties observed in real-world networks \cite{barabasi2016network}. For example, the configuration model preserves the exact degree (number of links incident to a node) of each node. Such models are used to understand how specific properties of networks affect dynamical processes on networks.

Recently, the computer science community has developed deep generative network models as a data-driven approach for generating realistic networks. Unlike classical network models in network science, which explicitly describe network formation mechanisms and preserve predefined topological properties, deep graph generative models use neural networks to learn the underlying probability distribution of graphs directly from network data \cite{zhu2022survey}. As a result, they can implicitly capture multiple network properties and their dependencies, generating realistic synthetic networks that closely resemble observed networks.
Deep graph generative models can augment limited graph datasets by generating additional realistic network instances, thereby expanding training sets for downstream graph learning tasks \cite{bonifati2021graph}. They can also facilitate network data sharing by generating synthetic graphs that preserve key statistical properties of the original networks while avoiding the release of individual-level connections, thereby reducing the risk of exposing sensitive relational information.

In this paper, we evaluate deep graph generative models from a network science perspective along two key dimensions: (1) topological similarity between real-world and model generated synthetic networks, and (2) their utility for a classic network science problem, namely epidemic immunization.
The epidemic immunization problem entails selecting a given number of nodes in a complex network to immunize in order to suppress the spread of epidemics or misinformation.
Progress has been made in designing immunization strategies for various epidemic spreading dynamics when the underlying social contact networks are given \cite{RevModPhys.87.925}.
However, sharing social networks, such as those collected in schools or local communities, raises privacy concerns due to the potential leakage of personal information. Rather than exposing real-world social networks, synthetic networks generated by graph models can be shared publicly to facilitate the development and evaluation of immunization strategies while preserving privacy.
Hence, we evaluate whether deep graph generative models can be used to identify effective immunization strategies. Four modern generative models based on autoregressive and diffusion approaches, along with the configuration model, will be considered to demonstrate the evaluation method.
We find that two generative models can effectively preserve multiple structural properties of real networks, allowing immunization strategies identified in synthetic networks to remain effective on real networks.

\section{Deep graph generative models}

We select four modern models that represent distinct modeling approaches and that we can scale to the neighborhood networks in our datasets within the memory limits of our evaluation hardware (notably a single GPU with 94 GB of memory).
Diffusion-like models (including related approaches such as flow-matching), followed by autoregressive ones, are most common in state-of-the-art literature and largely outperform approaches based on generative adversarial networks or variational autoencoders.
Correspondingly, our final selection includes one autoregressive model and three diffusion models.
Moreover, as a non-learning baseline, we also include the configuration model~\cite{barabasi2016network}.
It copies the degree sequence of a randomly selected real-world neighborhood network and randomly connects nodes, subject to the target degree sequence.

\textbf{GraphRNN}~\cite{graphrnn} represents a graph as a sequence and generates graphs sequentially, one node at a time.
A graph-level recurrent neural network (RNN) determines when to add or stop adding nodes, while a link-level RNN connects each new node to previous ones, using breadth-first search (BFS) orderings.
Its lightweight design makes GraphRNN fast and scalable.
However, its main limitations, including sensitivity to the processing order of nodes, difficulty in retaining long-range dependencies, and the inability to revise links once generated, are exacerbated as the graph size increases.

\textbf{EDGE}~\cite{edge} is a discrete diffusion model that operates directly on the adjacency matrix.
Generation starts from an empty graph of the desired size, after which the model repeatedly selects a subset of active nodes and predicts links to add between them.
Target node degrees, sampled from an auxiliary neural network or directly from the training data, help inform the selection and preserve connectivity statistics.
EDGE differs from more costly diffusion models by updating only a small active portion of the graph at each step, so it is particularly fast and scalable for sparse graphs.
However, its simpler neural network and diffusion process, which only allows adding links during generation, make it less expressive than other formulations.

\textbf{SparseDiff}~\cite{sparsediff} is also a discrete diffusion model emphasizing scalability to larger graphs.
Generation starts with a random graph whose link density matches the training set's empirical distribution.
Then, to form the final graph, the neural network chooses links to either add or remove over a predetermined number of iterations.
Structural and spectral features supplied at each step help the neural network achieve increased generation fidelity.
The key scaling mechanism is sparse, partial processing: SparseDiff stores links as adjacency lists and evaluates only a selected fraction of possible links at a time.
Compared to the original dense model it builds upon, SparseDiff trades generation speed for the ability to handle larger graphs.

\textbf{GGSD}~\cite{ggsd} applies diffusion in the spectral domain rather than directly to graph links.
It learns the continuous eigenvalues and eigenvectors of the graph Laplacian, as a proxy for reconstructing the adjacency matrix.
The spectral representation compactly captures important local and global structure without explicitly modeling every link.
GGSD includes two variants.
A partial-eigenpair variant diffuses a subset of the largest or smallest eigenpairs and uses a separate predictor to reconstruct the graph.
An all-eigenpair variant instead diffuses the entire spectrum and reconstructs the adjacency matrix without that predictor; it is more expensive but tends to improve generation quality.
GGSD's spectral representation is overall more computationally efficient than directly modeling the discrete link structure.

\section{Evaluation method}
\label{sec2}

Given a set of real-world networks, each graph generation model is used to generate a set of synthetic networks. The objective is to evaluate the similarity between these two sets of networks and to determine whether the set of synthetic networks could be used to identify effective immunization strategies.


\subsection{Real-world networks}
The following real-world networks will be considered.
\begin{enumerate}
  \item[*]Ia-email-EU \cite{leskovec2007graph}, an email contact network where nodes are individual email addresses, and two nodes are connected by an undirected link if they have any email communication.
  \item[*] CA-Astro Physics \cite{leskovec2007graph}: a co-authorship network constructed from papers submitted to the Astro Physics category of the arXiv e-print archive, where nodes represent authors and edges indicate co-authorship relationships.
  \item[*] Soc-advogato \cite{massa2009bowling}: a trust-based social network among users in the open-source software development community. A link is established between two nodes if they have certified each other.
  \item[*] Facebook \cite{leskovec2012learning}: a friendship network among users on Facebook.
  \item[*] CA-HepPh \cite{leskovec2007graph}: a co-authorship network constructed from papers at the arXiv High Energy Physics – Phenomenology (Hep–Ph) section.
\end{enumerate}
A set of networks will be sampled from each real-world network, representing local neighborhood networks. Each set of local community networks will be used as input for each network generation model to generate correspondingly 1000 synthetic networks.
Specifically, each real-world network is pre-processed by retaining the largest connected component, whose basic statistics are shown in the table \ref{tab:real_world_networks}; Furthermore, we sample a set of two-hop local neighborhood networks, each around a randomly selected node; Each neighborhood network consists of a randomly selected center node, its two-hop neighbors and all links between these nodes in the original network; We focus on medium-sized neighborhoods, i.e., those between the 20th and 80th percentiles of the node number distribution among the sampled local neighborhoods (between the 30th and 90th percentiles for CA-HepPh). These medium-sized neighborhood networks, numerous in number, have a high priority in immunization strategy development and will therefore be used to evaluate graph generative models. The number of medium-sized neighborhood networks sampled from each real-world networks is as follows: 6,773 for CA-HepPh, 10,792 for CA-AstroPh, 2,738 for Facebook, 19,604 for ia-email-EU, 3,042 for Soc-advogato.

\begin{table*}[t]
\centering
\caption{Basic statistics of each pre-processed real-world network $G$ considered:
the number of nodes ($N$), the number of links ($L$), the link density ($p$),
the average clustering coefficient ($C$), the diameter ($\rho$), and the network type.}
\label{tab:real_world_networks}
\small
\begin{tabular}{cccccccccccc}
\toprule
Network & $N$ & $L$ & $p$ & $C$ & $\rho$ & Type \\
\midrule
Ia-email-EU   & 32430 & 54397  & 0.0001  & 0.1127 & 9  & Email \\
Astro-Physics & 17903 & 197031 & 0.0012  & 0.6328 & 14 & Scientific collaboration \\
Soc-advogato  & 5054  & 41941  & 0.0033  & 0.2526 & 9  & Trust relationship \\
Facebook      & 4039  & 88234  & 0.0108  & 0.6055 & 8  & Social friendship \\
CA-HepPh      & 12008 & 118521 & 0.00164 & 0.6115 & 13 & Scientific collaboration \\
\bottomrule
\end{tabular}
\end{table*}

\subsection{Setup of graph generative models}
\label{sec:model-setup}

For all graph generative models, we adopt the training/generation hyperparameter configurations validated by the original authors on their largest evaluated graphs, making only adaptations needed for our local-neighborhood datasets.
For \textbf{GraphRNN}, we specify explicitly, before training, the largest size of the given set of neighborhood networks.
For \textbf{EDGE}, we choose to directly draw target node degrees from the observed training networks without involving an additional auxiliary network, ensuring that we work with degree sequences that exactly match the reference data.
In \textbf{SparseDiff}, we adopt the author's configuration for networks with community structure to represent more complex patterns, while considering only one-tenth of all possible links at a time to remain feasible on our larger networks.
Lastly, in \textbf{GGSD}, we use the complete graph spectrum, as we find it yields valid networks more consistently than partial representations across our datasets and remains computationally feasible.

\subsection{Epidemic immunization problem}
We use the classic Susceptible-Infected-Recovered (SIR) process to model epidemic/misinformation spreading on a network. In the SIR process, each node can be in only one of three possible states at any time: Susceptible (S), Infected (I), or Recovered (R). Initially, at $t=0$, a single seed node is infected, whereas the other nodes are susceptible. Each infected node independently infects each susceptible neighbor according to a Poisson process with infection rate $\beta$. Each infected node recovers (dead or permanently immune), also according to a Poisson process with rate $\gamma$.

We consider $\beta \in \{0.2,0.4,0.6\}$ and $\gamma=1$. This choice ensures that the epidemic spreads mostly in the real-world neighborhood networks considered, scenarios when epidemic mitigation is needed. Specifically,  we estimate the epidemic threshold $\tau_c$ of each neighborhood network using the heterogeneous mean-field approximation proposed by \cite{wang2016predicting}, as  $\tau_c = \frac{\langle k \rangle}{\langle k^2 \rangle - \langle k \rangle}$. Here, $\langle k \rangle$ and $\langle k^2 \rangle$ denote the first and second node degree moments, respectively.
Taking the CA-AstroPh dataset as an example, the estimated epidemic thresholds of most real local neighborhood networks are below 0.2. Hence, our choice of $\frac{\beta}{\gamma} \leq 0.2$ is above the epidemic threshold, ensuring a regime in which the outbreak persists rather than dying out.

The immunization problem asks how to select $f\%$ of nodes for immunization to minimize the average number of recovered nodes at the end (steady state) of the spreading process starting from a random seed node. The number of recovered nodes is also called outbreak size, i.e., the number of nodes that have ever been infected. Immunizing a node is equivalent to removing the node from the network at $t=0$.

\subsection{Immunization strategies}\label{Immunisation strategy and evaluation}

Nodal centrality metrics, which characterize various topological properties of nodes, have been widely used to identify target nodes for immunization because of their effectiveness and low computational complexity \cite{doostmohammadian2020centrality,dudkina2024comparison,Li2015Correlation,wang2008betweenness}. Each centrality metric could be used to rank nodes and nodes with the highest centrality values are selected for immunization. Hence, each centrality corresponds to a distinct immunization strategy.
We will consider a set of commonly used centrality metrics and evaluate whether the ranking of their effectiveness in mitigating epidemic spreading, obtained from the set of real neighborhood networks, is preserved when evaluated on the corresponding synthetic networks. In other words, we evaluate whether synthetic networks generated by deep graph generative models could identify the most effective strategy. The following centrality metrics will be considered.

\begin{enumerate}
  \item[*] Degree centrality $d_i$ of a node $i$ is the number of links incident to the node.

  \item[*] Leverage centrality $LC_i= \frac{1}{d_i} \sum_{j \in \mathcal{N}(i)} \frac{d_i - d_j}{d_i + d_j}$ of a node $i$ quantifies the relative difference in degree between node $i$ and its neighbors,
   where $\mathcal{N}(i)$ is the set of neighbors of node $i$.

  \item[*] Collective influence $\text{CI}^\ell_i=(d_i - 1) \sum_{j \in \partial \mathcal{B}(i, \ell)} (d_j - 1)$ of a node $i$ at radius $\ell$ captures both a node's local connectivity and its ability to infect distant nodes specified by radius $\ell$,  where $\partial \mathcal{B}(i, \ell)$ refers to the set of nodes located exactly at distance $\ell$ from $i$. We consider $\ell=2$, which is commonly used.

  \item[*] The k-shell centrality of a node quantifies its position within the core-periphery structure of a network.
  The k-core of a graph is obtained by recursively removing all nodes with degree less than k, together with their incident links, until every remaining node has degree at least k. The resulting subgraph is called the k-core. The k-shell centrality of a node is the largest value of k for which the node remains in the corresponding k-core.

  \item[*] The betweenness centrality of a node is the number of shortest paths between all pairs of nodes in the network that pass through the node.

  \item[*] The closeness centrality $c_i=\sum_{j \in V \backslash \{i\}}\frac{1}{{H_{i,j}}}$ of a node $i$ measures how close a node is connected to all the others via the shortest path, where $H_{i,j}$ is the hopcount of the shortest path between nodes $i$ and $j$.

  \item[*] The eigenvector centrality $x_i$ of node $i$ is the component of the principal eigenvector $x$ corresponding to node $ i$, and the principal eigenvector is the eigenvector corresponding to the largest eigenvalue $\lambda_1$ of the adjacency matrix $A$ of the static network. Hence, $x \lambda_1=Ax$.

  \item[*] The PageRank centrality $P_i=\frac{1 - \gamma}{N} + \gamma\sum_{j \in V \backslash \{i\}} \frac{A_{i,j}P_j}{k_{j}}$ of node $i$ is the probability that node $i$ is visited by a random walker.
  Here, $\gamma$ is the probability for a walker to move to a random neighbor of the current node being visited, while $1-\gamma$ is the probability for the walker to move to a random node. The $\gamma=0.85$ parameterization is a common choice for this centrality.
\end{enumerate}

\section{Performance analysis}\label{sec4}

In this section, we first analyze the similarity between real-world neighborhood networks and synthetic graphs generated by each model, and afterward assess each model's performance in identifying effective immunization strategies.
We consider the real-world network CA-AstroPh as a representative example to illustrate our findings. The main observations discussed below also hold for the other network datasets, which will be shared in the appendix \ref{Results of the other real-world networks}.

\subsection{Network similarity}
\label{Synthetic networks quality evaluation}

\begin{figure*}[t]
    \centering
    \subfloat[Node number]{%
        \includegraphics[width=0.33\textwidth]{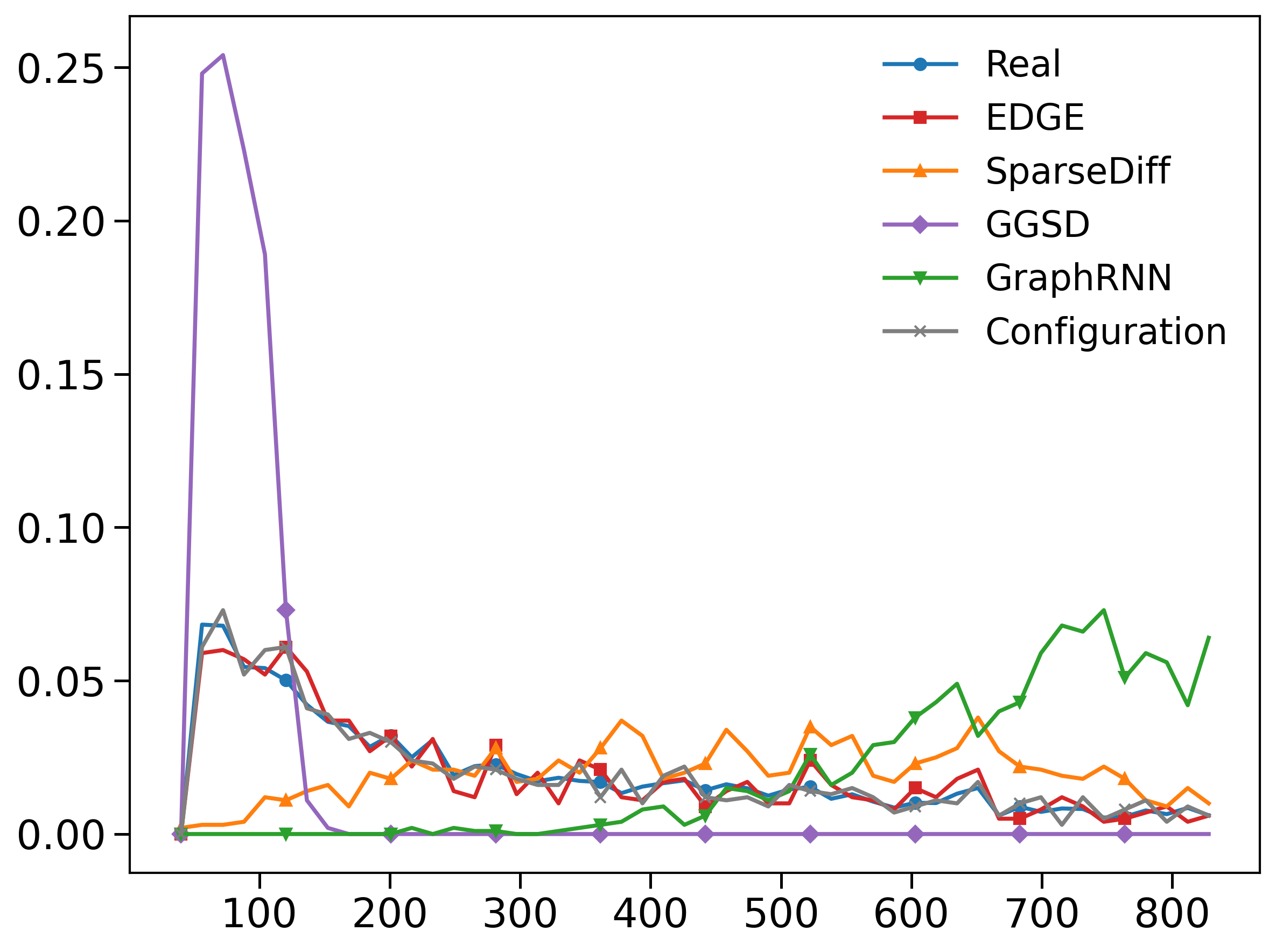}}
    \hfill
    \subfloat[Node degrees]{%
        \includegraphics[width=0.33\textwidth]{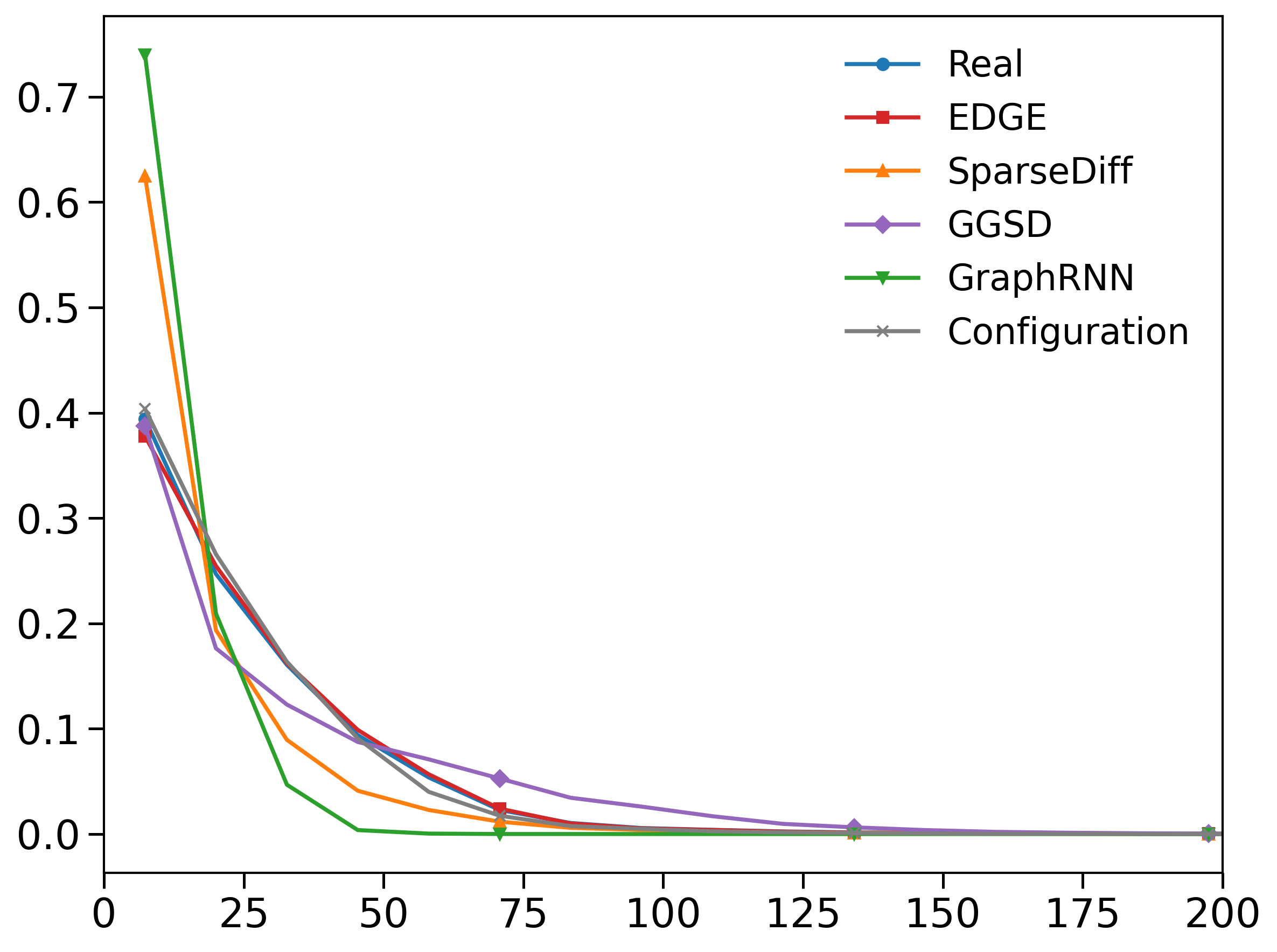}}
    \hfill
    \subfloat[Clustering coefficient]{%
        \includegraphics[width=0.33\textwidth]{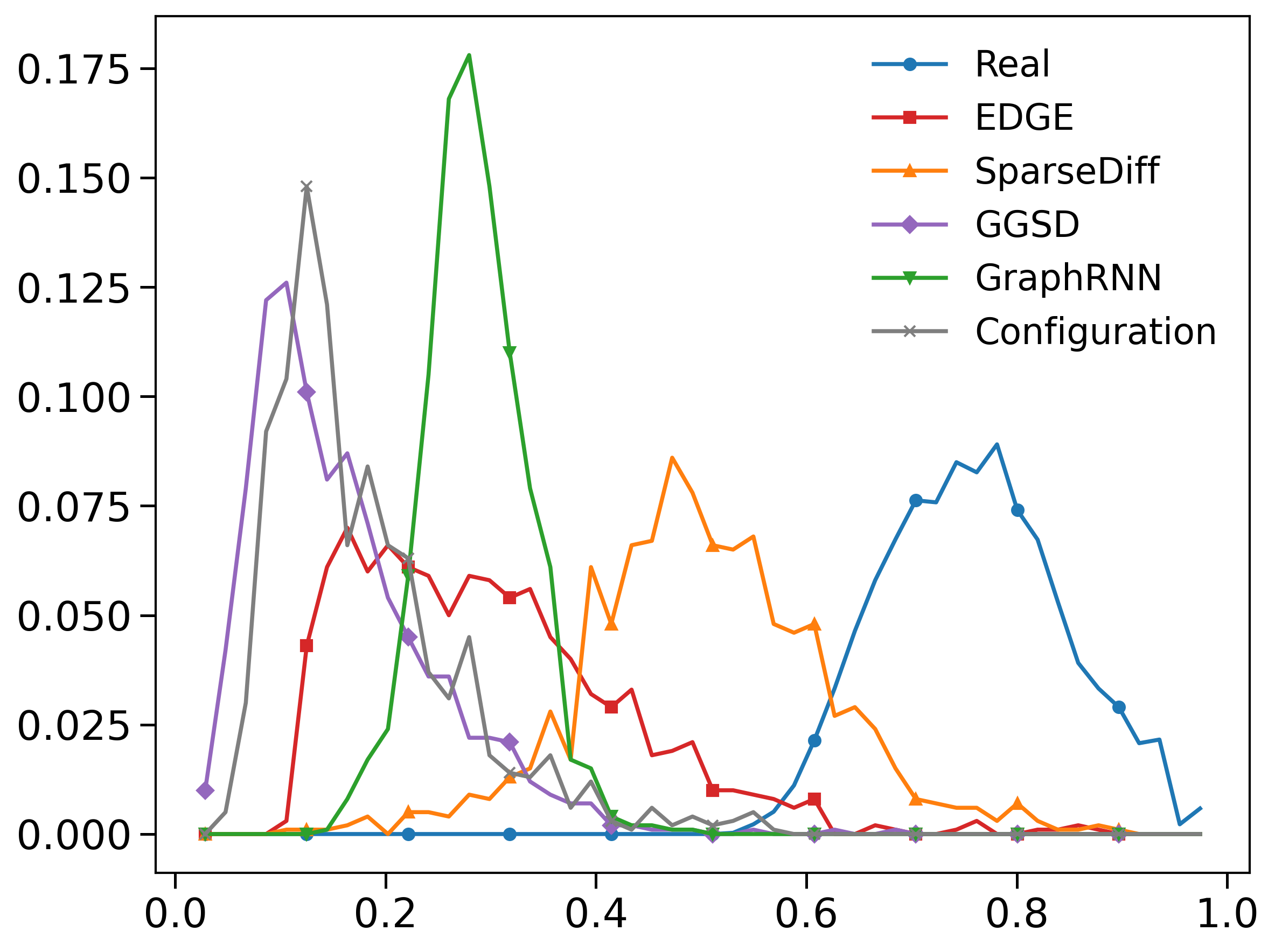}}

    \vspace{6pt}

    \subfloat[Largest eigenvalue]{%
        \includegraphics[width=0.33\textwidth]{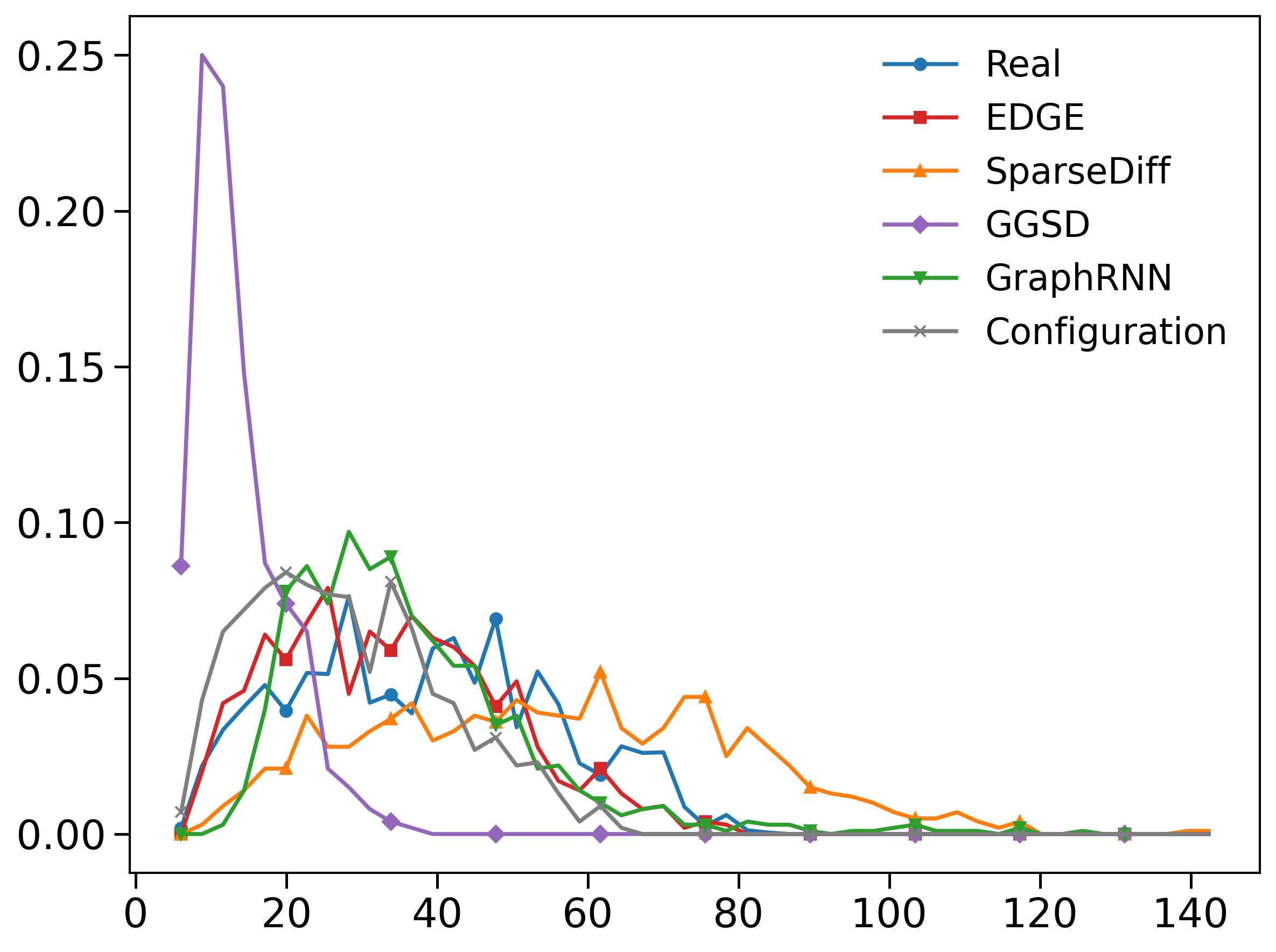}}
    \hfill
    \subfloat[Link density]{%
        \includegraphics[width=0.33\textwidth]{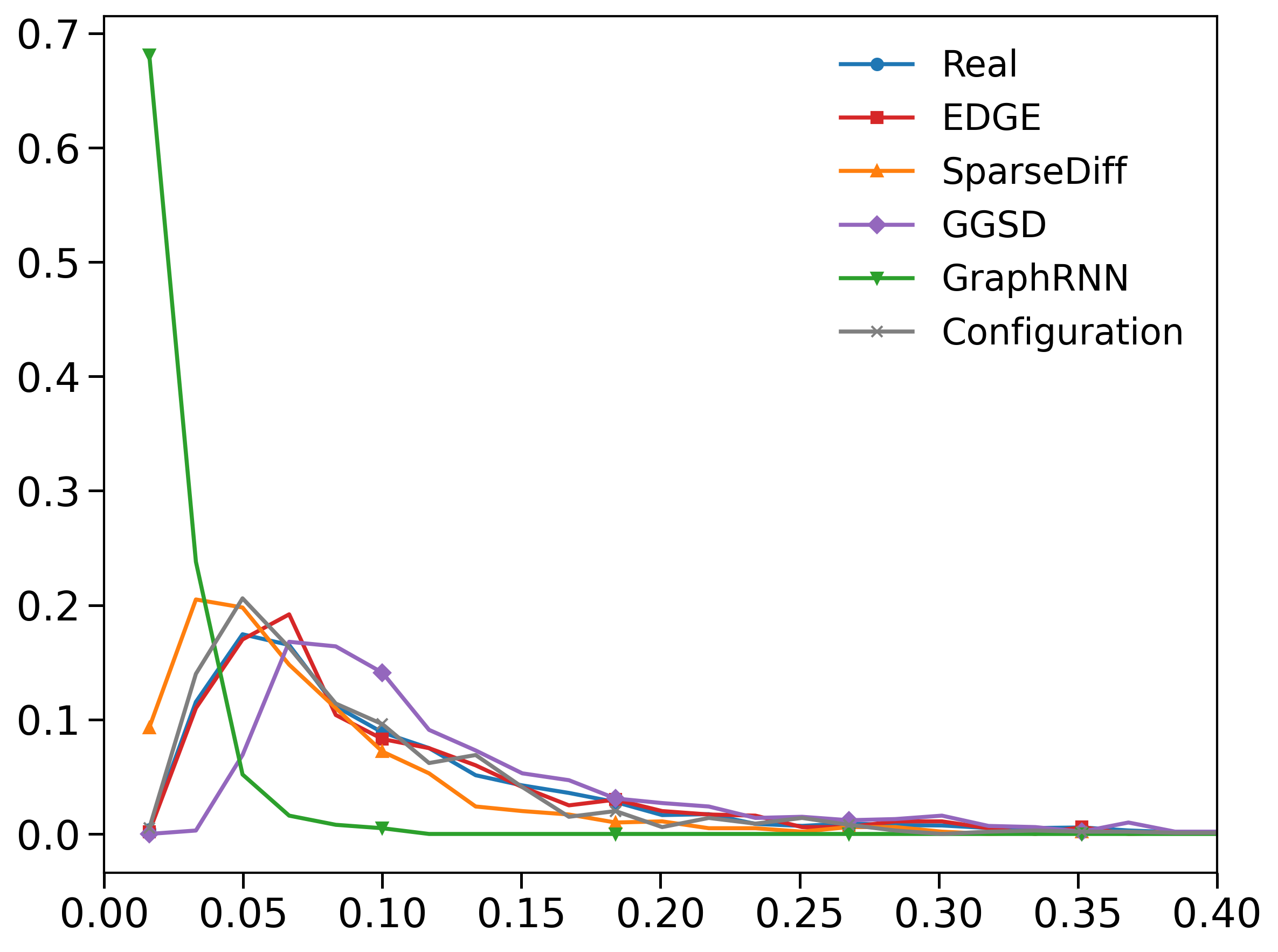}}
        \hfill
    \subfloat[Modularity]{%
        \includegraphics[width=0.33\textwidth]{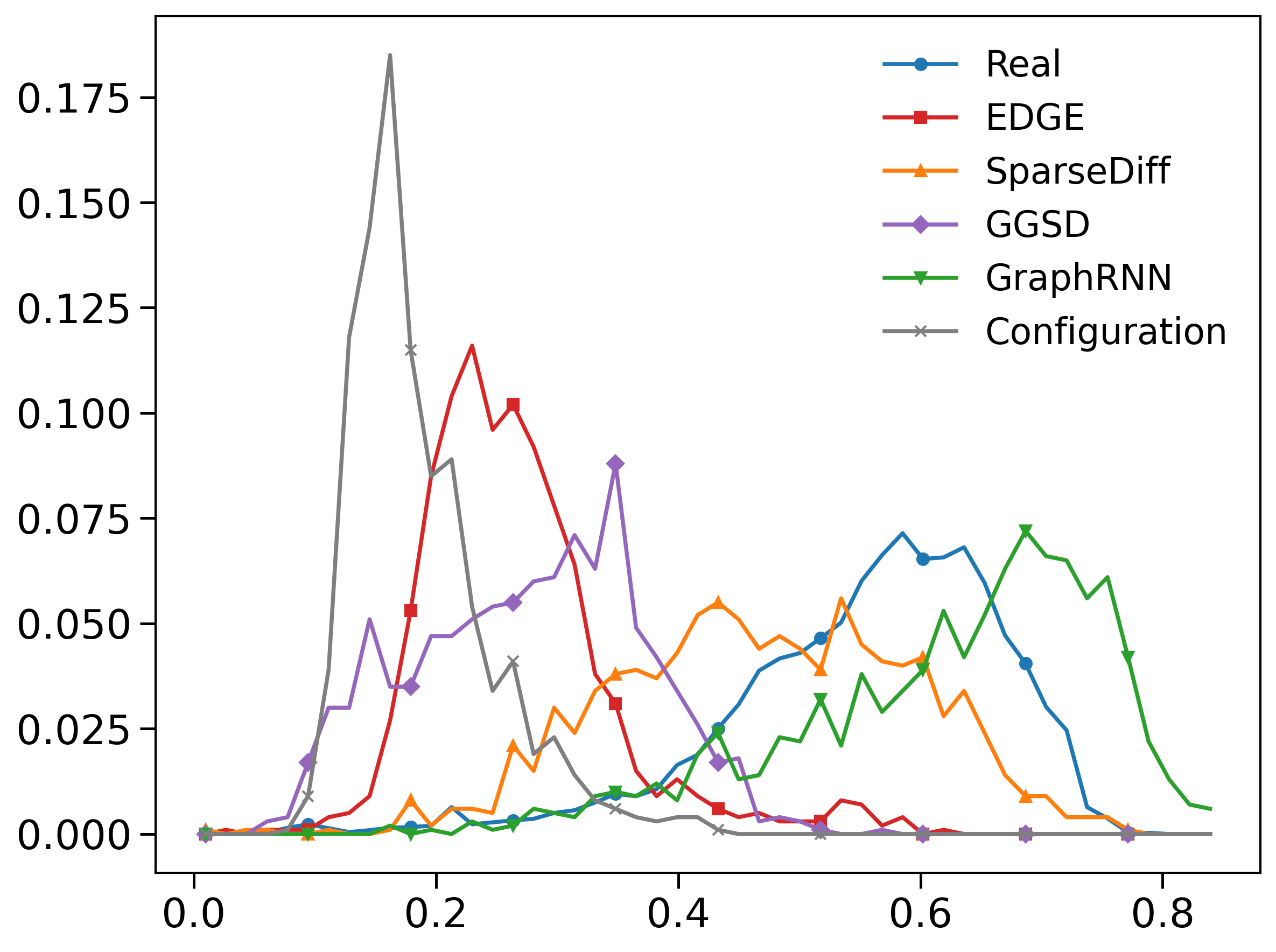}}
    \caption{Probability distributions of fundamental network properties, including the node number, node degree, clustering coefficient, largest eigenvalue, link density, and modularity for real-world neighborhood networks derived from CA-AstroPh dataset and the corresponding synthetic networks generated by each model.}
    \label{fig:CA-AstroPh-distribution}
\end{figure*}

We evaluate the topological similarity between synthetic and real neighborhood networks by comparing the distributions of the following graph- and node-level properties: node count, node degree, link density (the fraction of present links relative to all possible edges), nodal clustering coefficient (local density among neighbors), modularity (community structure strength relative to a random baseline) and the largest eigenvalue of the adjacency matrix.
\Cref{fig:CA-AstroPh-distribution} compares the distributions of these properties for synthetic and real-world neighborhood networks sampled from CA-AstroPh.

No single model universally excels; rather, each reproduces specific network topological properties more accurately, demonstrating that model selection should be tailored to the target application. However, notable discrepancies remain between real-world networks and the synthetic networks generated by the Configuration model, GraphRNN, and GGSD. As expected, the configuration model generates random graphs with low clustering coefficients and low modularity, while preserving the original degree distribution and network size. Consequently, it fails to capture both macroscopic properties such as modularity and microscopic property clustering coefficients. GGSD generates networks that are evidently smaller and have a more heterogeneous degree distribution. This also explains why networks generated by GGSD have a far smaller largest eigenvalue and a smaller clustering coefficient. GraphRNN generates far larger networks with evidently lower link density. This suggests that these synthetic networks tend to be disconnected into small components, supported by their high modularity. The other two models, Edge and SparseDiff, reproduce properties of real-world networks more accurately.



\subsection{Evaluating immunization strategies}

In this subsection, we evaluate which network generation model performs better in identifying effective immunization strategies and estimating their effect.
Given a real-world neighborhood network, or a corresponding synthesized network, given the infection rate $\beta \in \{0.2, 0.4, 0.6\}$ and recover rate $\gamma=1$, the average outbreak size is obtained for each strategy to immunize $fN$ nodes, over all possible single seed node and $200$ realizations of the spreading process starting from each seed node. The effect of an immunization strategy, evaluated on a set of real-world neighborhood networks sampled from a real-world network (or on the corresponding set of synthetic networks generated by a model), is quantified by the average outbreak size, then averaged over all networks in that set.

\begin{figure*}[t]
    \centering
    \includegraphics[width=1.0\textwidth]{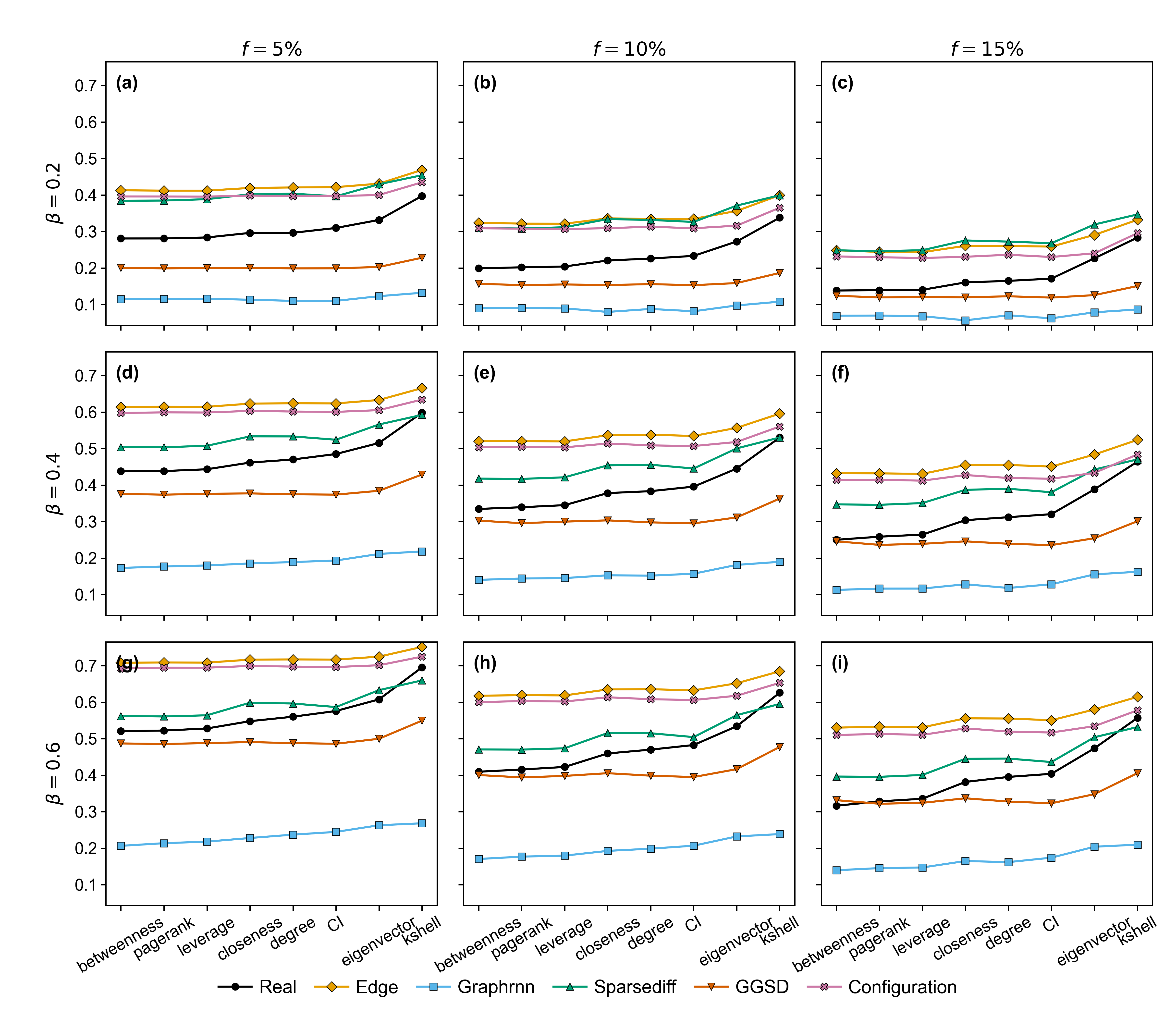}
    \caption{
        Performances of considered immunization strategies across infection rates
        $\beta \in \{0.2,0.4,0.6\}$, evaluated by outbreak size on the
        CA-AstroPh dataset, for immunization rates
        $f \in \{5\%,10\%,15\%\}$.
    }
    \label{fig:CA-AstroPh-immunization}
\end{figure*}

\Cref{fig:CA-AstroPh-immunization} shows the performance of the eight strategies on the set of real-world neighborhoods as well as on the set of networks synthesized by each model. On real-world networks, betweenness, Pagerank, and leverage centrality consistently perform best, whereas k-shell centrality performs worst.
This can be partially explained by examining the network properties of the pruned networks after immunized nodes are removed. A network tends to be robust against epidemic spreading if it has a small largest eigenvalue, a small largest connected component (LCC), and a high modularity \cite{schneider2011mitigation,PhysRevE.88.022801,zhang2022mitigate}. Table \ref{tab:property_change} shows that, indeed, the pruned networks resulting from the three best-performing strategies tend to have a smaller largest eigenvalue, a smaller largest connected component, and higher modularity than those resulting from the least effective strategy, K-shell.

Among all models, the SparseDiff model generally performs best in two respects. First, it most accurately reproduces the ranking of immunization strategies observed on real-world networks: it correctly identifies the three most effective strategies and the two least effective ones. Second, it most accurately reproduces the average outbreak size associated with each strategy. The EDGE model also perform well in accurately identifying the most and worst performance immunization strategies. The out-performance of SparseDiff and EDGE models has also been observed in other considered datasets. This finding is consistent with the observation in \ref{Synthetic networks quality evaluation} that these two models generate synthetic networks that are more similar to real-world neighborhood networks. The configuration model precisely reproduces the degree distribution of real networks but cannot preserve more complex structural properties, such as clustering coefficient and modularity. Therefore, although the degree distribution is an important factor influencing the SIR process \cite{RevModPhys.87.925}, the configuration model still fails to predict the ranking of immunization strategies on real-world networks accurately.

\begin{table}[htbp]
\centering
\caption{The average of modularity, the largest eigenvalue, and the LCC size of pruned real-world networks under different immunization strategies, based on the CA-AstroPh dataset, when the immunization rate $f=10\%$.}
\label{tab:property_change}
\resizebox{\linewidth}{!}{%
\begin{tabular}{cccccccccccc}
\hline
 & Betweenness & PageRank & Leverage & Closeness & Degree & Collective Influence & Eigenvector & K-shell \\
\hline
Modularity
& 0.6504
& 0.6598
& 0.6576
& 0.6602
& 0.6931
& 0.5547
& 0.6433
& 0.5808 \\

Eigenvalue
& 29.6374
& 28.3577
& 27.9576
& 28.5799
& 23.4232
& 34.6438
& 25.9660
& 28.9021 \\

LCC size
& 240.3363
& 242.4618
& 244.6125
& 248.5461
& 251.0587
& 264.1254
& 256.4318
& 263.2287 \\
\hline
\end{tabular}
}
\end{table}

\section{Conclusion and future work}\label{sec5}

In this paper, we systematically evaluate deep graph generative models from a network science perspective by assessing both the topological similarity between synthetic networks they generate and real-world networks and their utility in identifying effective immunization strategies. It is found that that the SparseDiff and EDGE models can generate synthetic networks that relatively closely preserve multiple structural properties of real-world networks and, consequently, enable the identification of effective immunization strategies. These findings suggest that network generation models could facilitate the development of network-based applications in scenarios where real-world network data cannot be shared.

We consider only one basic immunization problem, in which the spreading process is modeled by the SIR process, to illustrate our evaluation method. Given the promise of deep graph generative models in identifying effective immunization strategies in this basic problem setup, future work should explore (1) more realistic spreading models (2) diverse network models beyond the configure model \cite{vaca2022systematic,vu2025ec} and (3) systematically evaluating privacy preservation across models.
Finally, each graph generative model requires a set of real-world networks, in our case, neighborhood networks, for training. When the goal is to identify immunization strategies for larger neighborhood networks, the scalability of graph generation models should be considered in greater detail.

\section*{Acknowledgements}
We thank for the support of this work from Netherlands Organisation for Scientific Research NWO (
project FORT-PORT no. KICH1.VE03.21.008, and NEPTARGOS no. KICH1.VE05.23.003) and NExTWORKx, a collaboration between TUDelft and KPN on future telecommunication networks.
\bibliographystyle{spmpsci} 
\bibliography{refs} 

@inproceedings{zhu2022survey,
  title={A survey on deep graph generation: Methods and applications},
  author={Zhu, Yanqiao and Du, Yuanqi and Wang, Yinkai and Xu, Yichen and Zhang, Jieyu and Liu, Qiang and Wu, Shu},
  booktitle={Learning on Graphs Conference},
  pages={47--1},
  year={2022},
  organization={PMLR}
}

@article{bonifati2021graph,
  title={Graph generators: State of the art and open challenges},
  author={Bonifati, Angela and Holubov{\'a}, Irena and Prat-P{\'e}rez, Arnau and Sakr, Sherif},
  journal={ACM Computing Surveys},
  volume={53},
  number={2},
  year={2021}
}

@inproceedings{graphrnn,
  author       = {Jiaxuan You and
                  Rex Ying and
                  Xiang Ren and
                  William L. Hamilton and
                  Jure Leskovec},
  title        = {GraphRNN: Generating Realistic Graphs with Deep Auto-regressive Models},
  booktitle    = {{ICML}},
  series       = {Proceedings of Machine Learning Research},
  volume       = {80},
  pages        = {5694--5703},
  publisher    = {{PMLR}},
  year         = {2018}
}

@inproceedings{edge,
  author       = {Xiaohui Chen and
                  Jiaxing He and
                  Xu Han and
                  Liping Liu},
  title        = {Efficient and Degree-Guided Graph Generation via Discrete Diffusion
                  Modeling},
  booktitle    = {{ICML}},
  series       = {Proceedings of Machine Learning Research},
  volume       = {202},
  pages        = {4585--4610},
  publisher    = {{PMLR}},
  year         = {2023}
}

@article{Sparsediff,
  author       = {Yiming Qin and
                  Cl{\'{e}}ment Vignac and
                  Pascal Frossard},
  title        = {SparseDiff: Sparse Discrete Diffusion for Scalable Graph Generation},
  journal      = {Trans. Mach. Learn. Res.},
  volume       = {2025},
  year         = {2025}
}

@inproceedings{ggsd,
  author       = {Giorgia Minello and
                  Alessandro Bicciato and
                  Luca Rossi and
                  Andrea Torsello and
                  Luca Cosmo},
  title        = {Generating Graphs via Spectral Diffusion},
  booktitle    = {{ICLR}},
  publisher    = {OpenReview.net},
  year         = {2025}
}

@article{doostmohammadian2020centrality,
  title={Centrality-based epidemic control in complex social networks},
  author={Doostmohammadian, Mohammadreza and Rabiee, Hamid R and Khan, Usman A},
  journal={Social Network Analysis and Mining},
  volume={10},
  pages={1--11},
  year={2020},
  publisher={Springer}
}

@article{dudkina2024comparison,
  title={A comparison of centrality measures and their role in controlling the spread in epidemic networks},
  author={Dudkina, Ekaterina and Bin, Michelangelo and Breen, Jane and Crisostomi, Emanuele and Ferraro, Pietro and Kirkland, Steve and Mare{\v{c}}ek, Jakub and Murray-Smith, Roderick and Parisini, Thomas and Stone, Lewi and others},
  journal={International Journal of Control},
  volume={97},
  number={6},
  pages={1325--1340},
  year={2024},
  publisher={Taylor \& Francis}
}

@article{wang2016predicting,
  title={Predicting the epidemic threshold of the susceptible-infected-recovered model},
  author={Wang, Wei and Liu, Quan-Hui and Zhong, Lin-Feng and Tang, Ming and Gao, Hui and Stanley, H Eugene},
  journal={Scientific reports},
  volume={6},
  number={1},
  pages={24676},
  year={2016},
  publisher={Nature Publishing Group UK London}
}

@article{leskovec2007graph,
     title={Graph evolution: Densification and shrinking diameters},
     author={Leskovec, Jure and Kleinberg, Jon and Faloutsos, Christos},
     journal={ACM Transactions on Knowledge Discovery from Data (TKDD)},
     volume={1},
     number={1},
     pages={2},
     year={2007},
     publisher={ACM}
}

@article{wang2008betweenness,
  title={Betweenness centrality in a weighted network},
  author={Wang, Huijuan and Hernandez, Javier Martin and Van Mieghem, Piet},
  journal={Physical Review E},
  volume={77},
  number={4},
  pages={046105},
  year={2008},
  publisher={APS}
}

@inproceedings{massa2009bowling,
  title={Bowling alone and trust decline in social network sites},
  author={Massa, Paolo and Salvetti, Martino and Tomasoni, Danilo},
  booktitle={2009 Eighth IEEE International Conference on Dependable, Autonomic and Secure Computing},
  pages={658--663},
  year={2009},
  organization={IEEE}
}

@article{zhang2022mitigate,
  title={Mitigate SIR epidemic spreading via contact blocking in temporal networks},
  author={Zhang, Shilun and Zhao, Xunyi and Wang, Huijuan},
  journal={Applied network science},
  volume={7},
  number={1},
  pages={2},
  year={2022},
  publisher={Springer}
}

@article{leskovec2012learning,
  title={Learning to discover social circles in ego networks},
  author={Leskovec, Jure and Mcauley, Julian},
  journal={Advances in neural information processing systems},
  volume={25},
  year={2012}
}

@article{schneider2011mitigation,
  title={Mitigation of malicious attacks on networks},
  author={Schneider, Christian M and Moreira, Andr{\'e} A and Andrade Jr, Jos{\'e} S and Havlin, Shlomo and Herrmann, Hans J},
  journal={Proceedings of the National Academy of Sciences},
  volume={108},
  number={10},
  pages={3838--3841},
  year={2011},
  publisher={National Academy of Sciences}
}

@article{Li2015Correlation,
  author    = {Cong Li and Qian Li and Piet Van Mieghem and H. Eugene Stanley and Huijuan Wang},
  title     = {Correlation between centrality metrics and their application to the opinion model},
  journal   = {The European Physical Journal B},
  volume    = {88},
  number    = {3},
  pages     = {65},
  year      = {2015}
}

@article{PhysRevE.88.022801,
  title = {Effect of the interconnected network structure on the epidemic threshold},
  author = {Wang, Huijuan and Li, Qian and D'Agostino, Gregorio and Havlin, Shlomo and Stanley, H. Eugene and Van Mieghem, Piet},
  journal = {Phys. Rev. E},
  volume = {88},
  issue = {2},
  pages = {022801},
  numpages = {13},
  year = {2013},
  month = {Aug},
  publisher = {American Physical Society},
}

@book{barabasi2016network,
  author = {Barabási, Albert-László and Pósfai, Márton},
  isbn = {9781107076266 1107076269},
  publisher = {Cambridge University Press},
  title = {Network science},
  year = {2016}
}

@article{RevModPhys.87.925,
  title = {Epidemic processes in complex networks},
  author = {Pastor-Satorras, Romualdo and Castellano, Claudio and Van Mieghem, Piet and Vespignani, Alessandro},
  journal = {Rev. Mod. Phys.},
  volume = {87},
  issue = {3},
  pages = {925--979},
  numpages = {55},
  year = {2015},
  month = {Aug},
  publisher = {American Physical Society},
}

@article{vaca2022systematic,
  title={Systematic assessment of the quality of fit of the stochastic block model for empirical networks},
  author={Vaca-Ram{\'\i}rez, Felipe and Peixoto, Tiago P},
  journal={Physical Review E},
  volume={105},
  number={5},
  pages={054311},
  year={2022},
  publisher={APS}
}

@article{vu2025ec,
  title={EC-SBM synthetic network generator},
  author={Vu-Le, The-Anh and Anne, Lahari and Chacko, George and Warnow, Tandy},
  journal={Applied Network Science},
  volume={10},
  number={1},
  pages={15},
  year={2025},
  publisher={Springer}
}

\appendix
\section{Results of the other real-world networks}
\label{Results of the other real-world networks}

\begin{figure*}[t]
    \centering
    \subfloat[Node number]{%
        \includegraphics[width=0.33\textwidth]{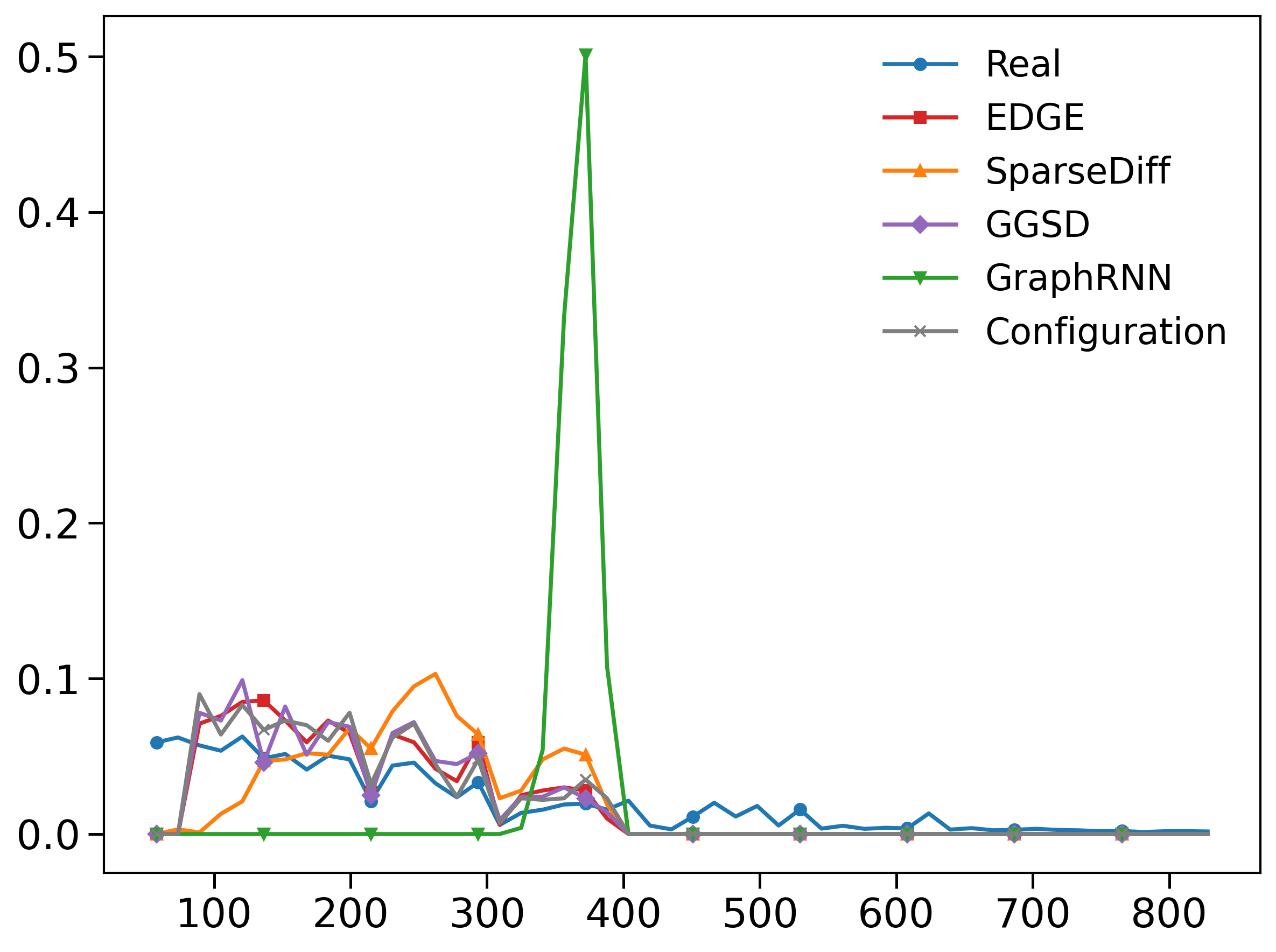}}
    \hfill
    \subfloat[Node degrees]{%
        \includegraphics[width=0.33\textwidth]{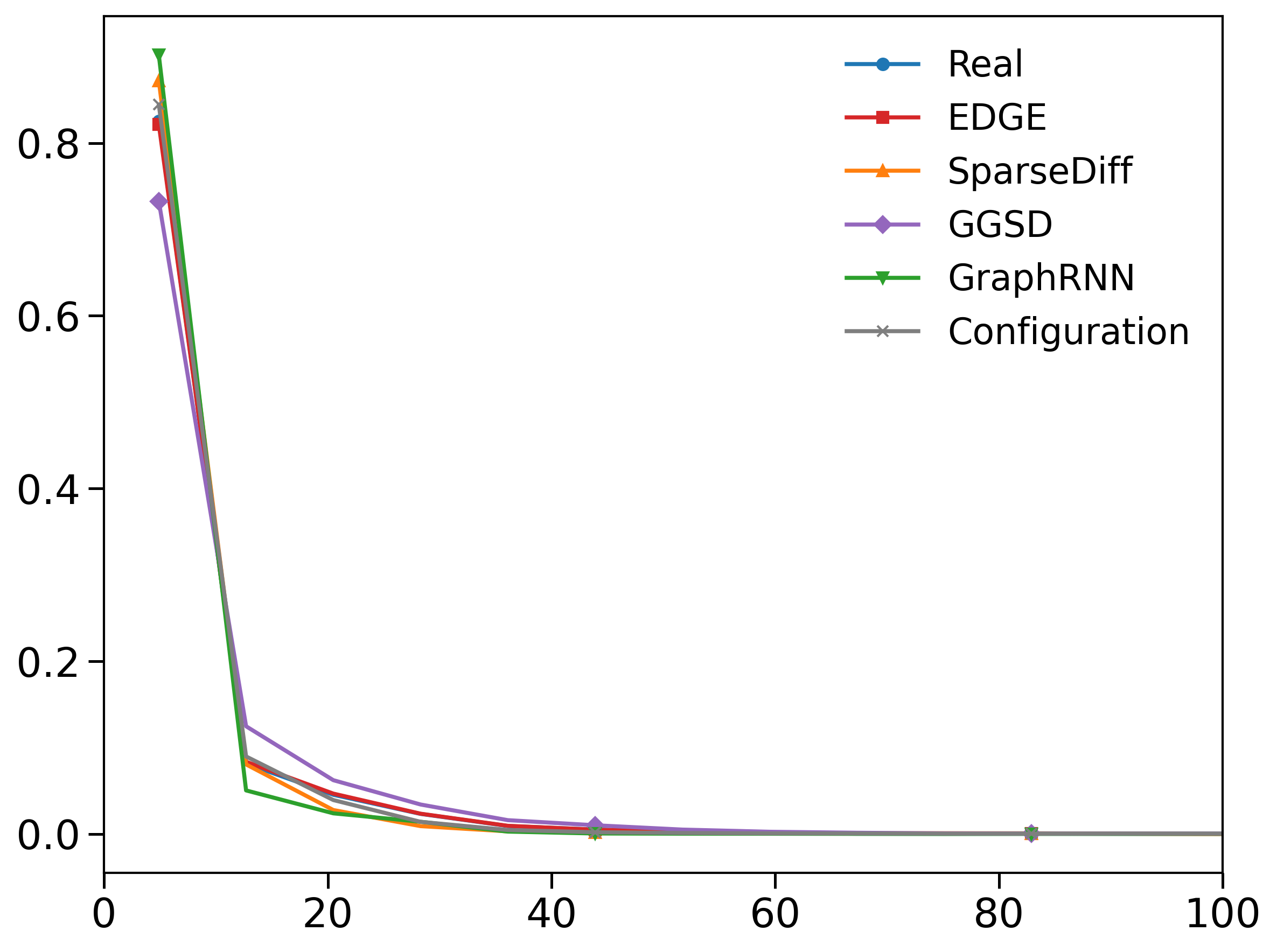}}
    \hfill
    \subfloat[Clustering coefficient]{%
        \includegraphics[width=0.33\textwidth]{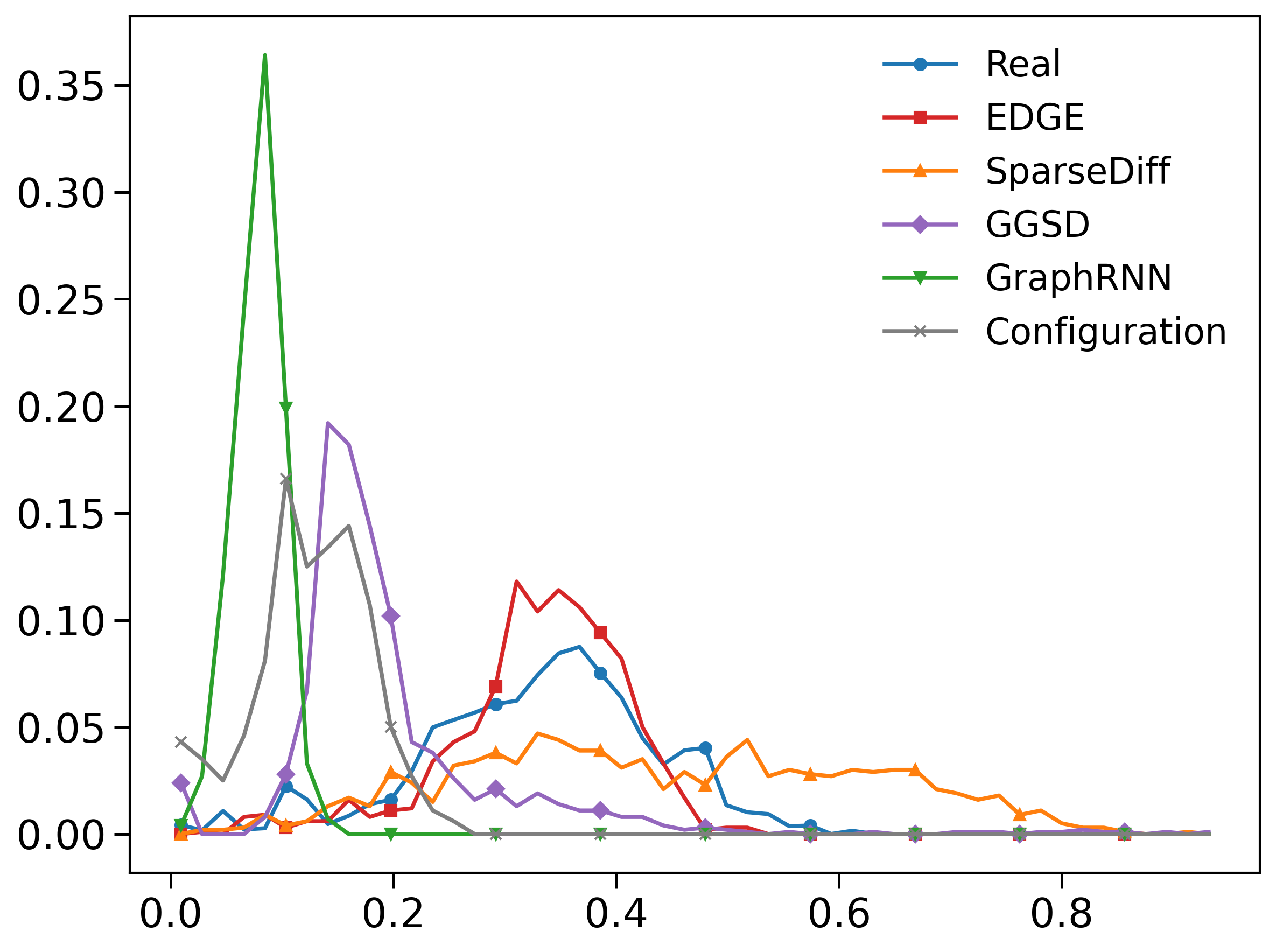}}

    \vspace{6pt}

    \subfloat[Largest eigenvalue]{%
        \includegraphics[width=0.33\textwidth]{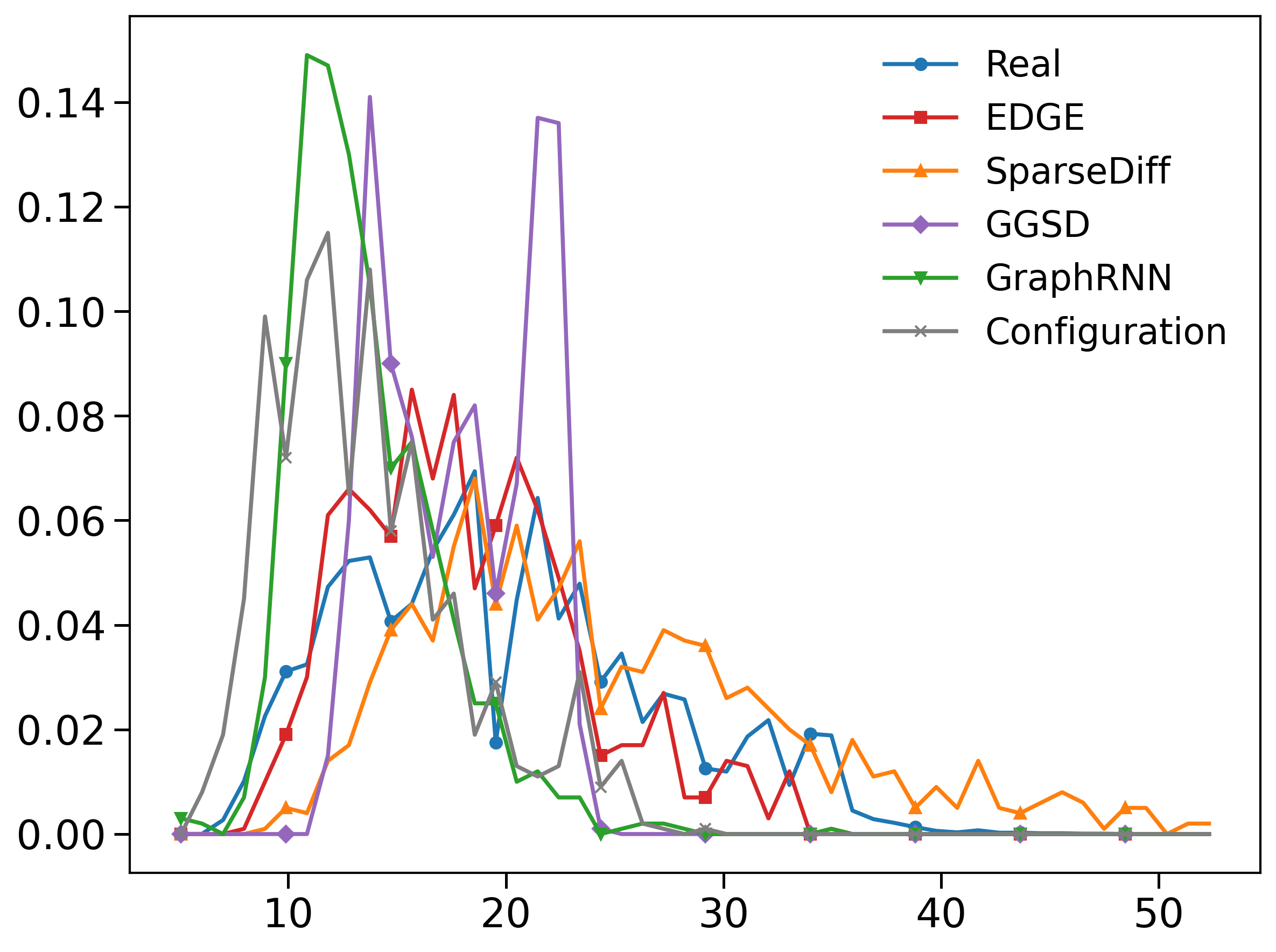}}
    \hfill
    \subfloat[Link density]{%
        \includegraphics[width=0.33\textwidth]{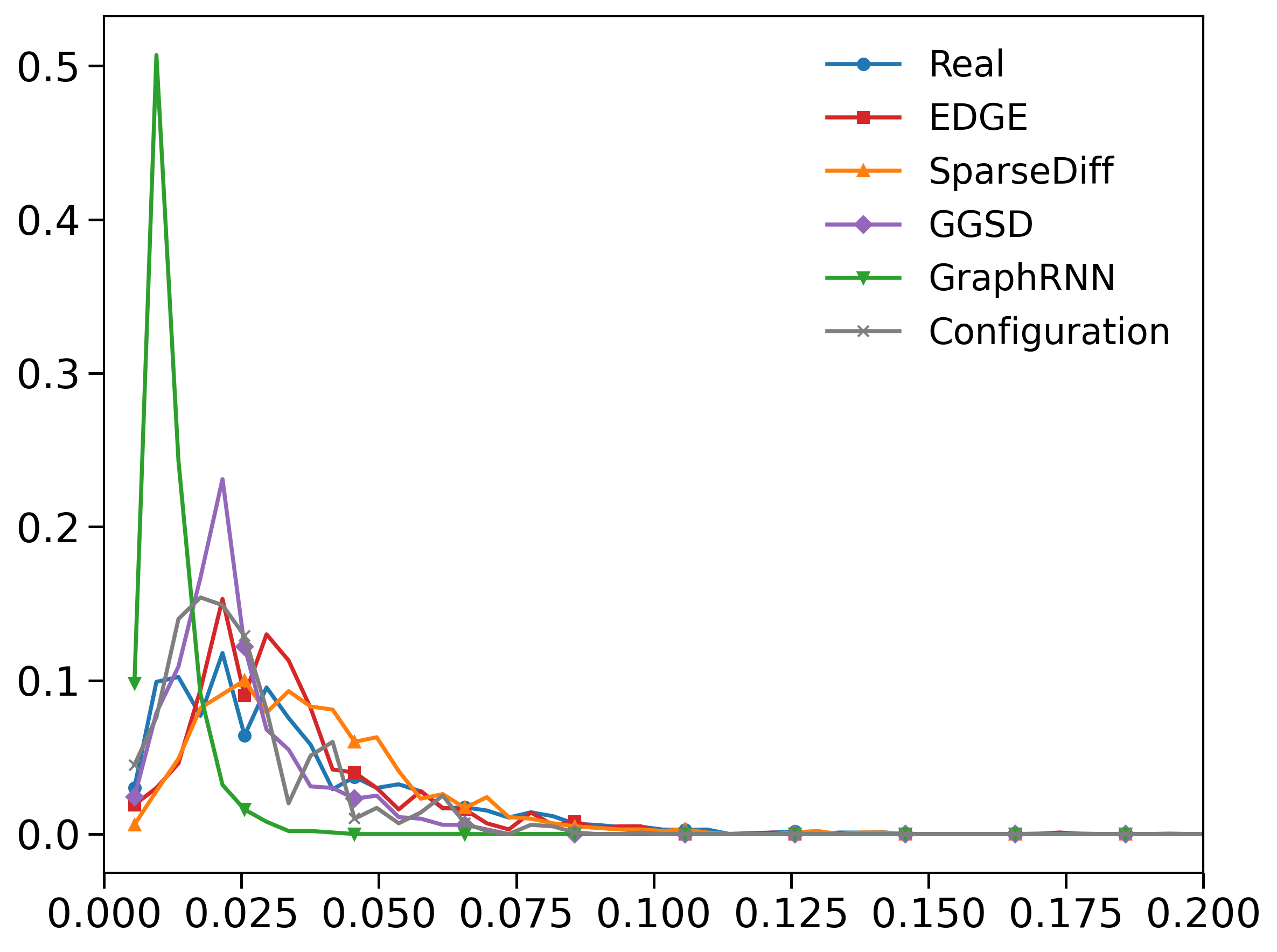}}
        \hfill
    \subfloat[Modularity]{%
        \includegraphics[width=0.33\textwidth]{Subnetworks_statistics/CA-AstroPh_modularity.png}}
    \caption{Probability distributions of fundamental network properties, including the node number, node degree, clustering coefficient, largest eigenvalue, link density, and modularity for real-world neighborhood networks derived from ia-email-EU dataset and the corresponding synthetic networks generated by each model.}
    \label{fig:ia-email-EU-distribution}
\end{figure*}

\begin{figure*}[t]
    \centering
    \includegraphics[width=1.0\textwidth]{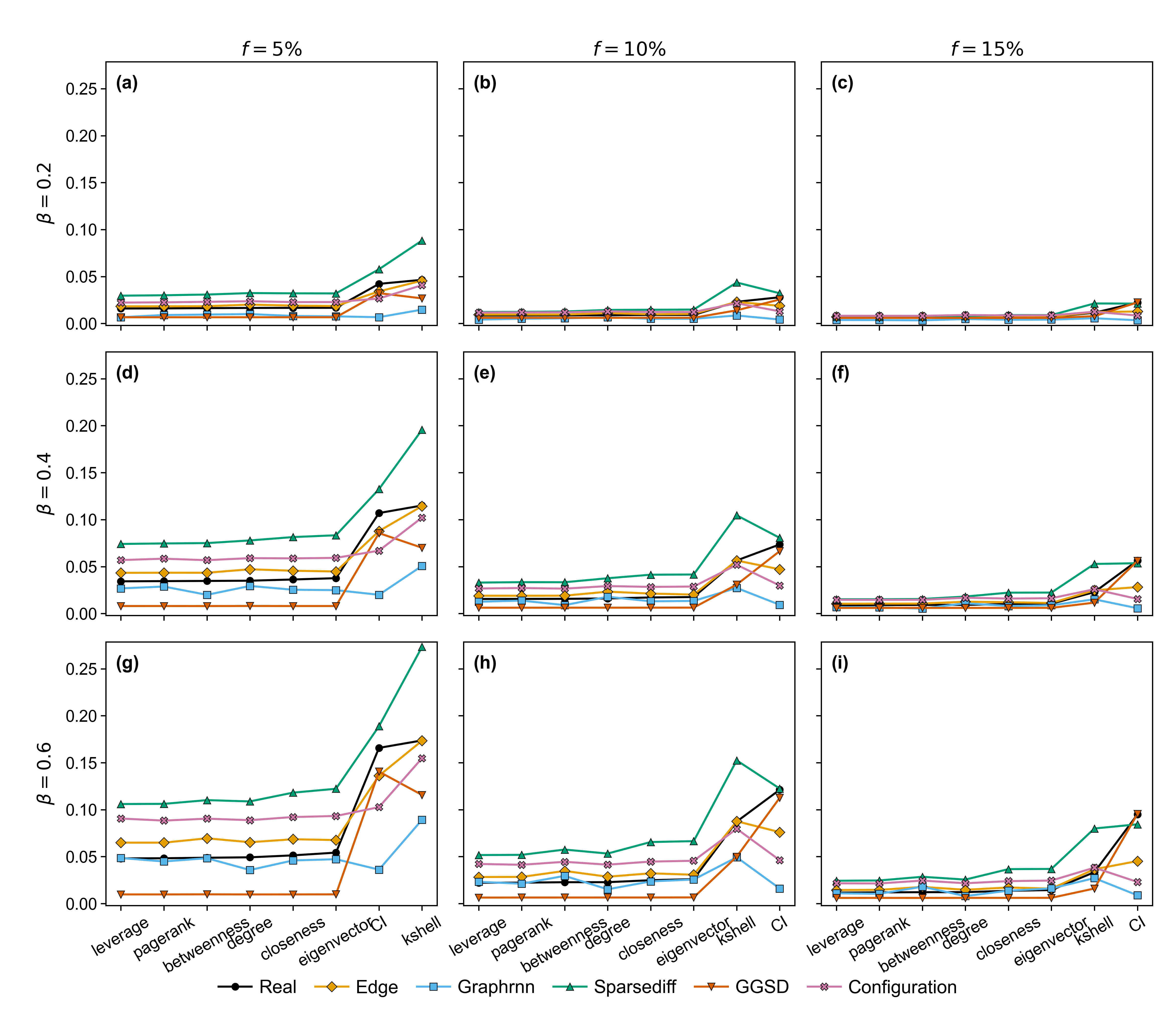}
    \caption{
        Performances of considered immunization strategies across infection rates
        $\beta \in \{0.2,0.4,0.6\}$, evaluated by outbreak size on the
        ia-email-EU dataset, for immunization rates
        $f \in \{5\%,10\%,15\%\}$.
    }
    \label{fig:ia-email-EU-immunization}
\end{figure*}

\begin{figure*}[t]
    \centering
    \subfloat[Node number]{%
        \includegraphics[width=0.33\textwidth]{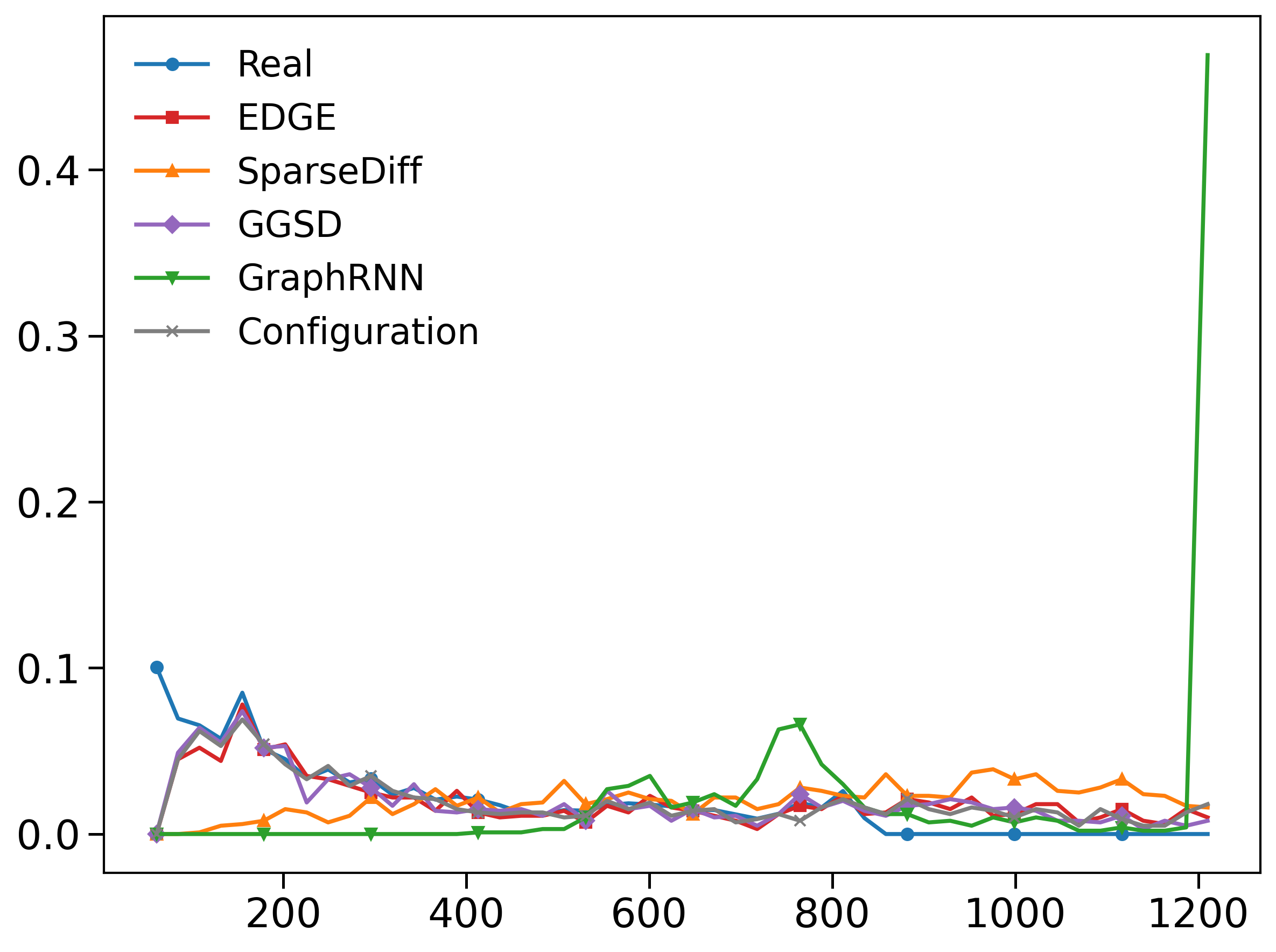}}
    \hfill
    \subfloat[Node degrees]{%
        \includegraphics[width=0.33\textwidth]{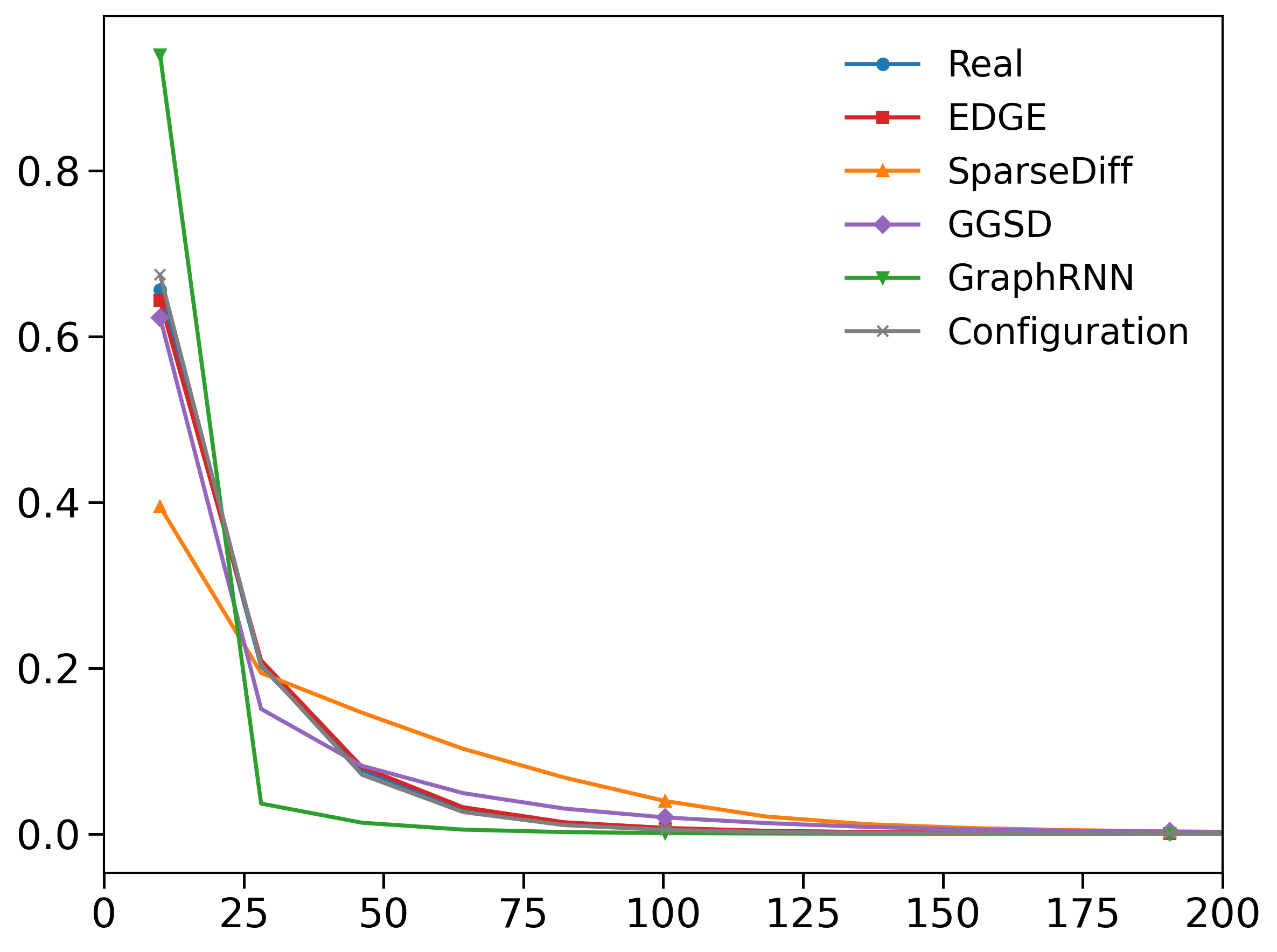}}
    \hfill
    \subfloat[Clustering coefficient]{%
        \includegraphics[width=0.33\textwidth]{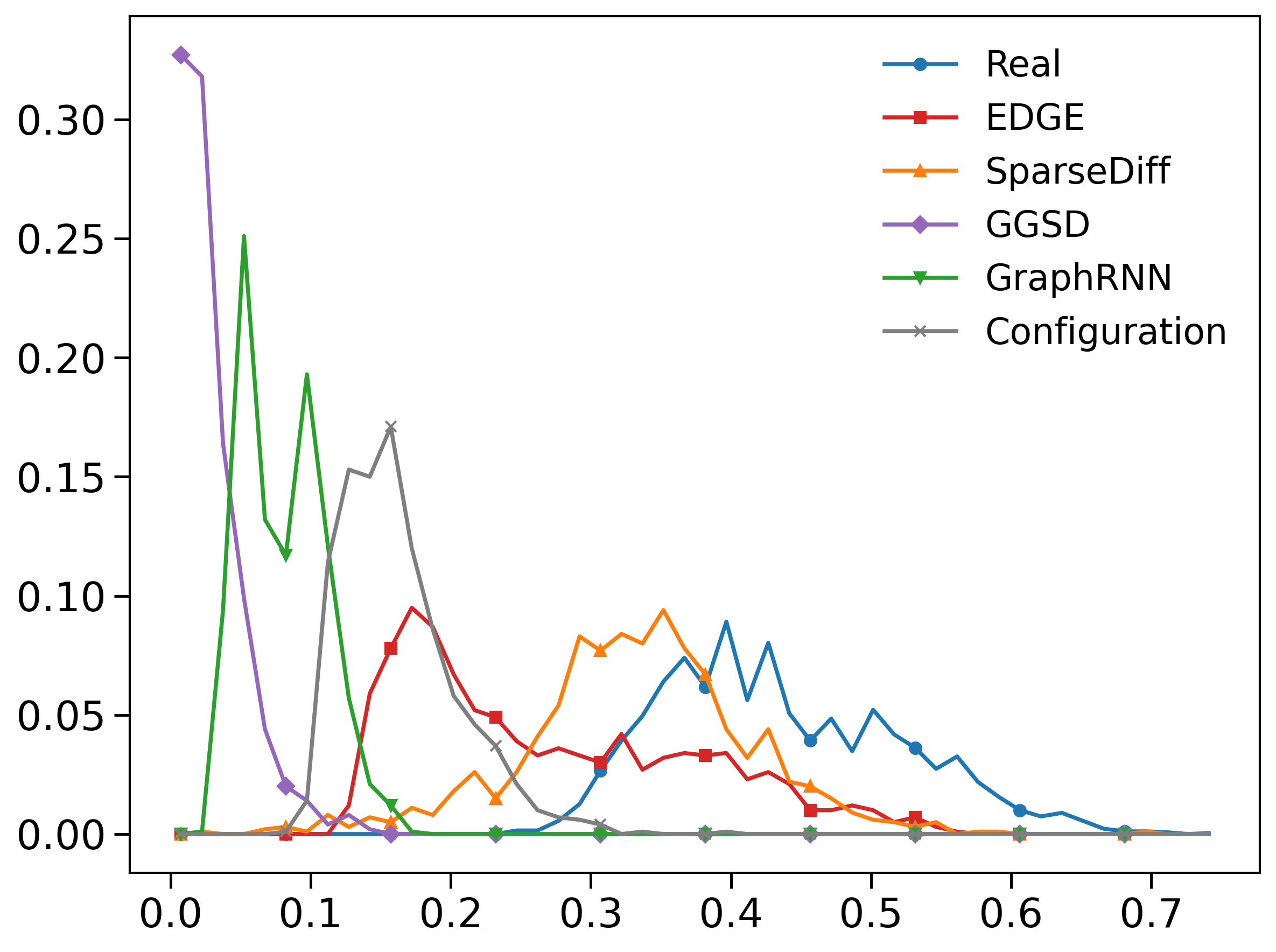}}

    \vspace{6pt}

    \subfloat[Largest eigenvalue]{%
        \includegraphics[width=0.33\textwidth]{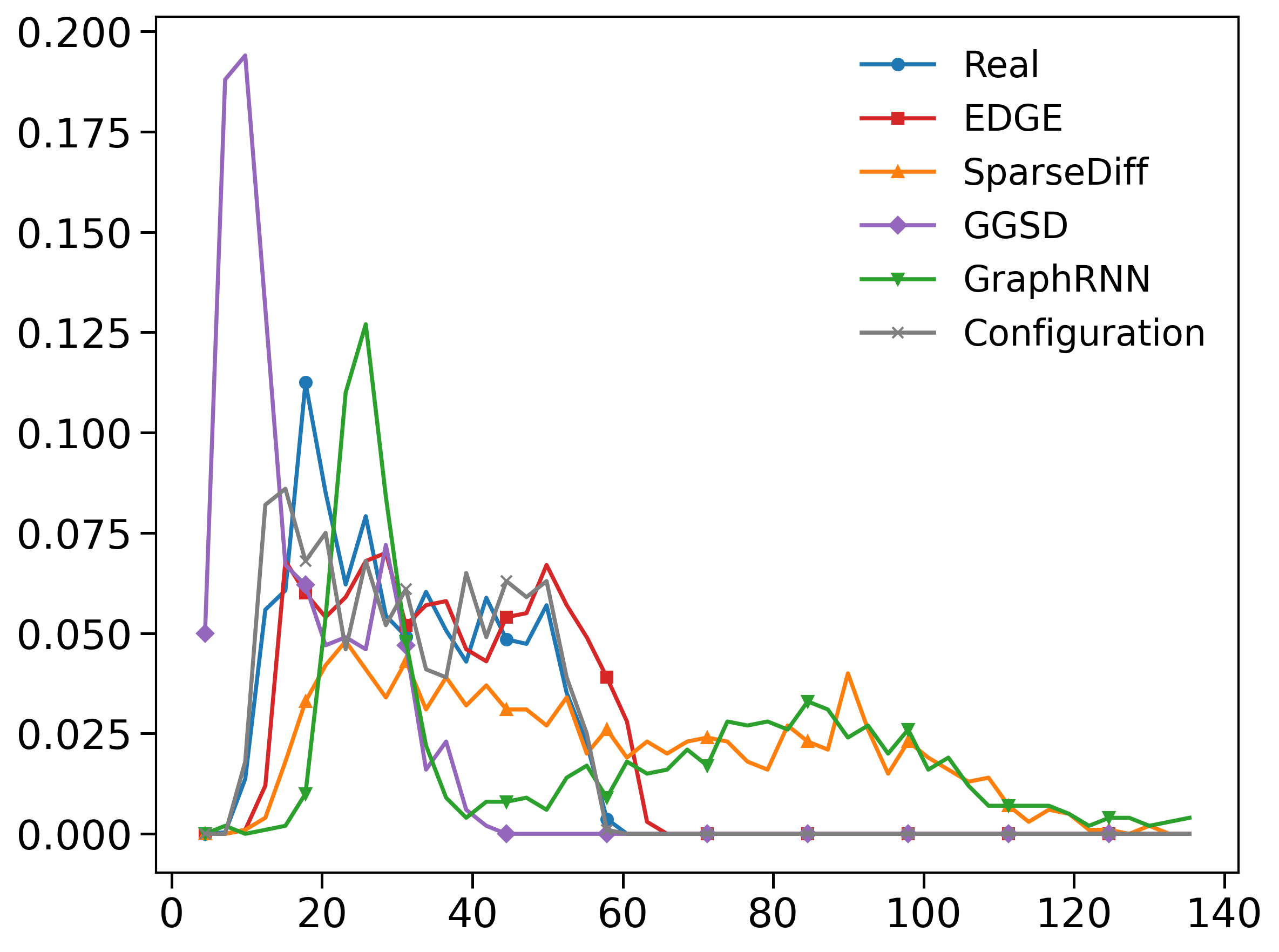}}
    \hfill
    \subfloat[Link density]{%
        \includegraphics[width=0.33\textwidth]{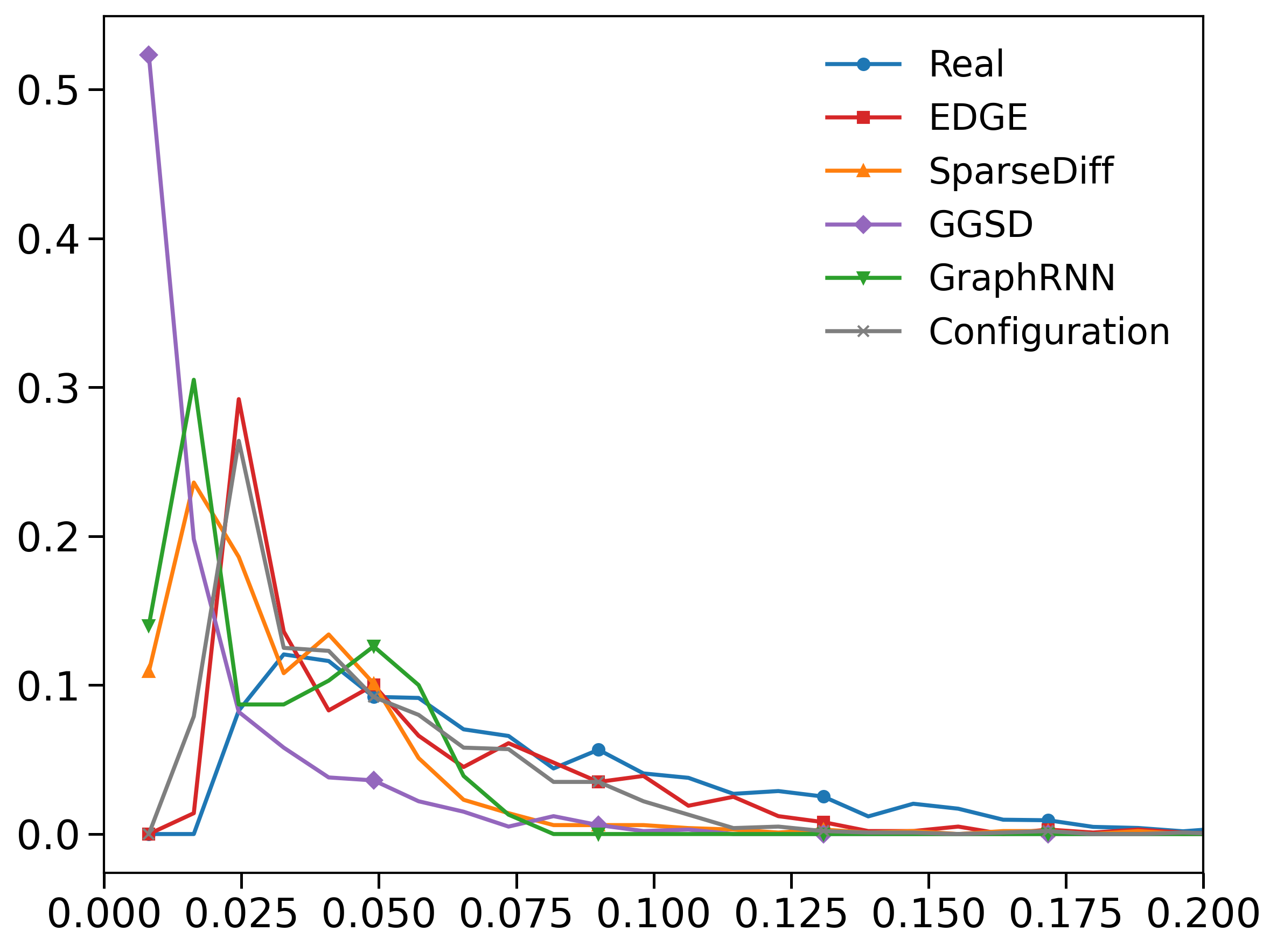}}
        \hfill
    \subfloat[Modularity]{%
        \includegraphics[width=0.33\textwidth]{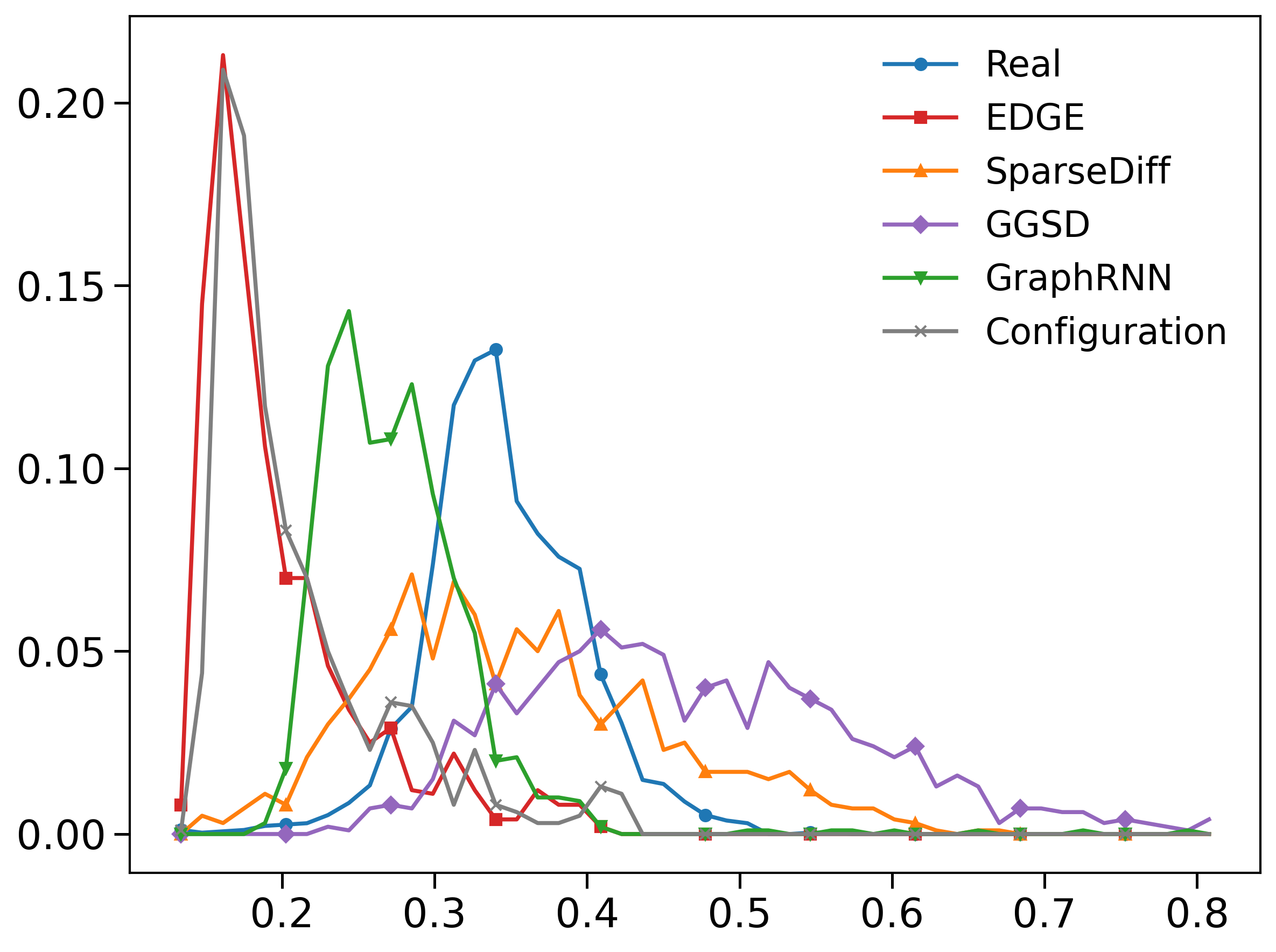}}
    \caption{Probability distributions of fundamental network properties, including the node number, node degree, clustering coefficient, largest eigenvalue, link density, and modularity for real-world neighborhood networks derived from soc-advogato dataset and the corresponding synthetic networks generated by each model.}
    \label{fig:soc-advogato-distribution}
\end{figure*}

\begin{figure*}[t]
    \centering
    \includegraphics[width=1.0\textwidth]{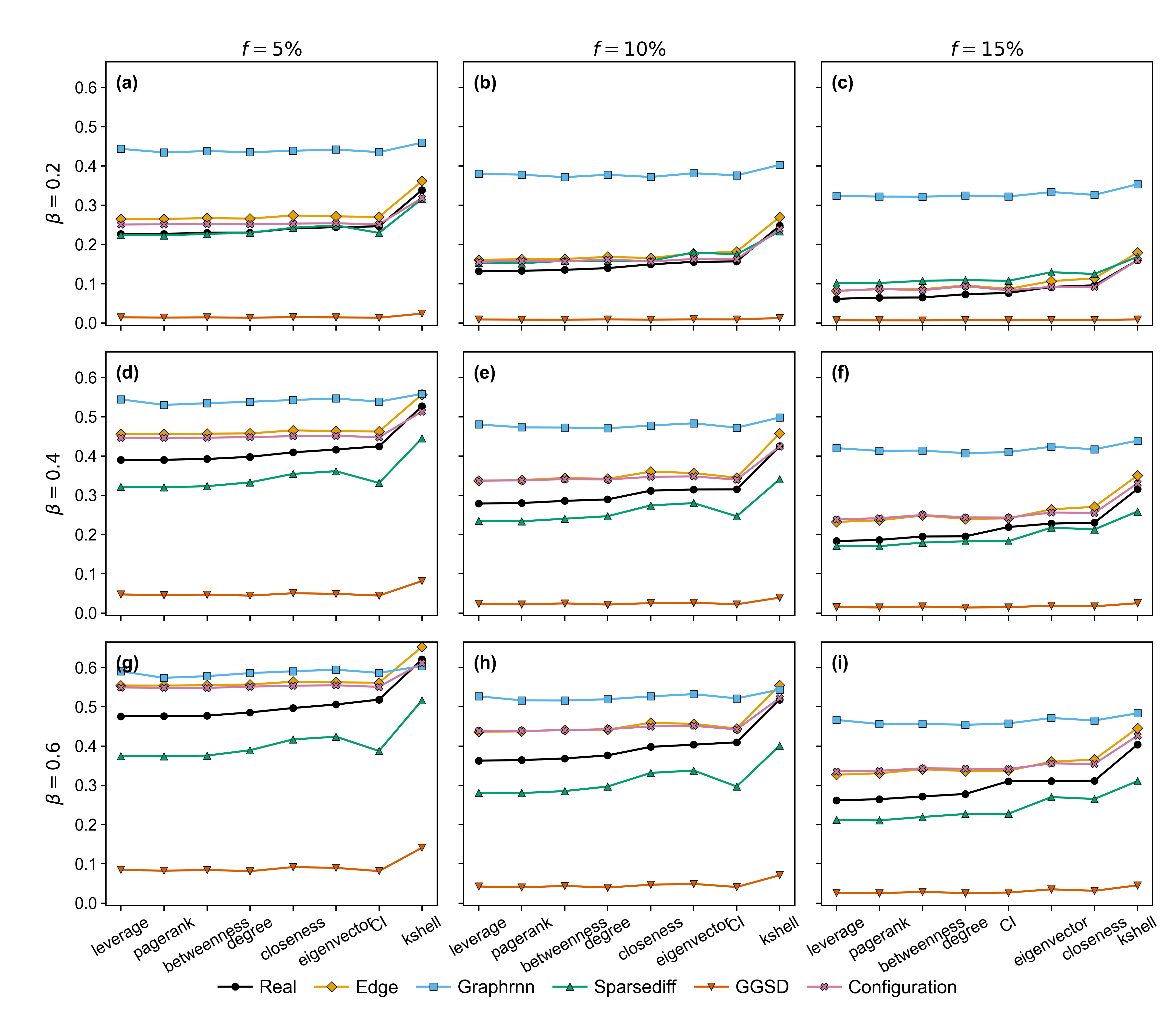}
    \caption{
        Performances of considered immunization strategies across infection rates
        $\beta \in \{0.2,0.4,0.6\}$, evaluated by outbreak size on the
        soc-advogato dataset, for immunization rates
        $f \in \{5\%,10\%,15\%\}$.
    }
    \label{fig:soc-advogato-immunization}
\end{figure*}

\begin{figure*}[t]
    \centering
    \subfloat[Node number]{%
        \includegraphics[width=0.33\textwidth]{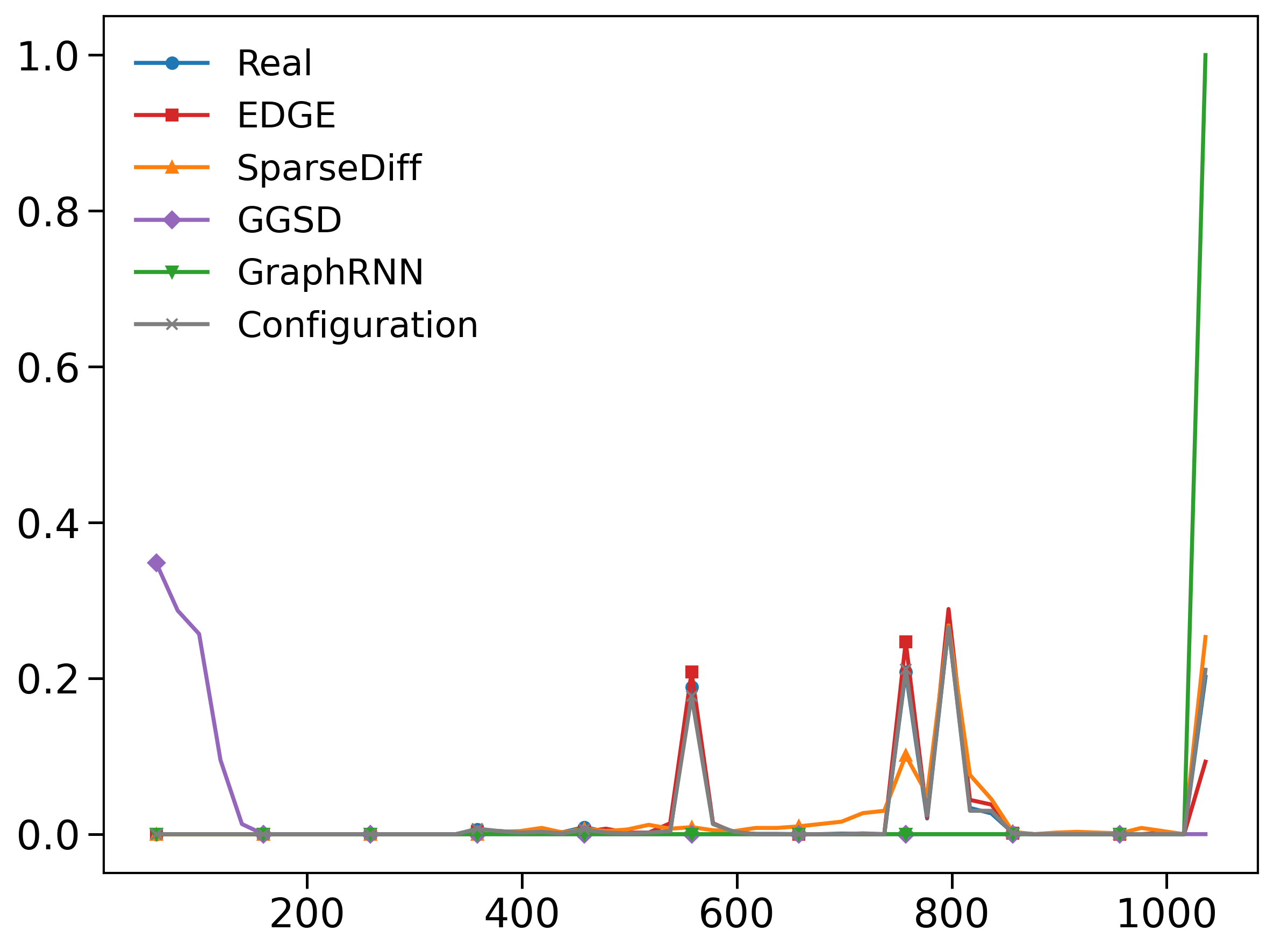}}
    \hfill
    \subfloat[Node degrees]{%
        \includegraphics[width=0.33\textwidth]{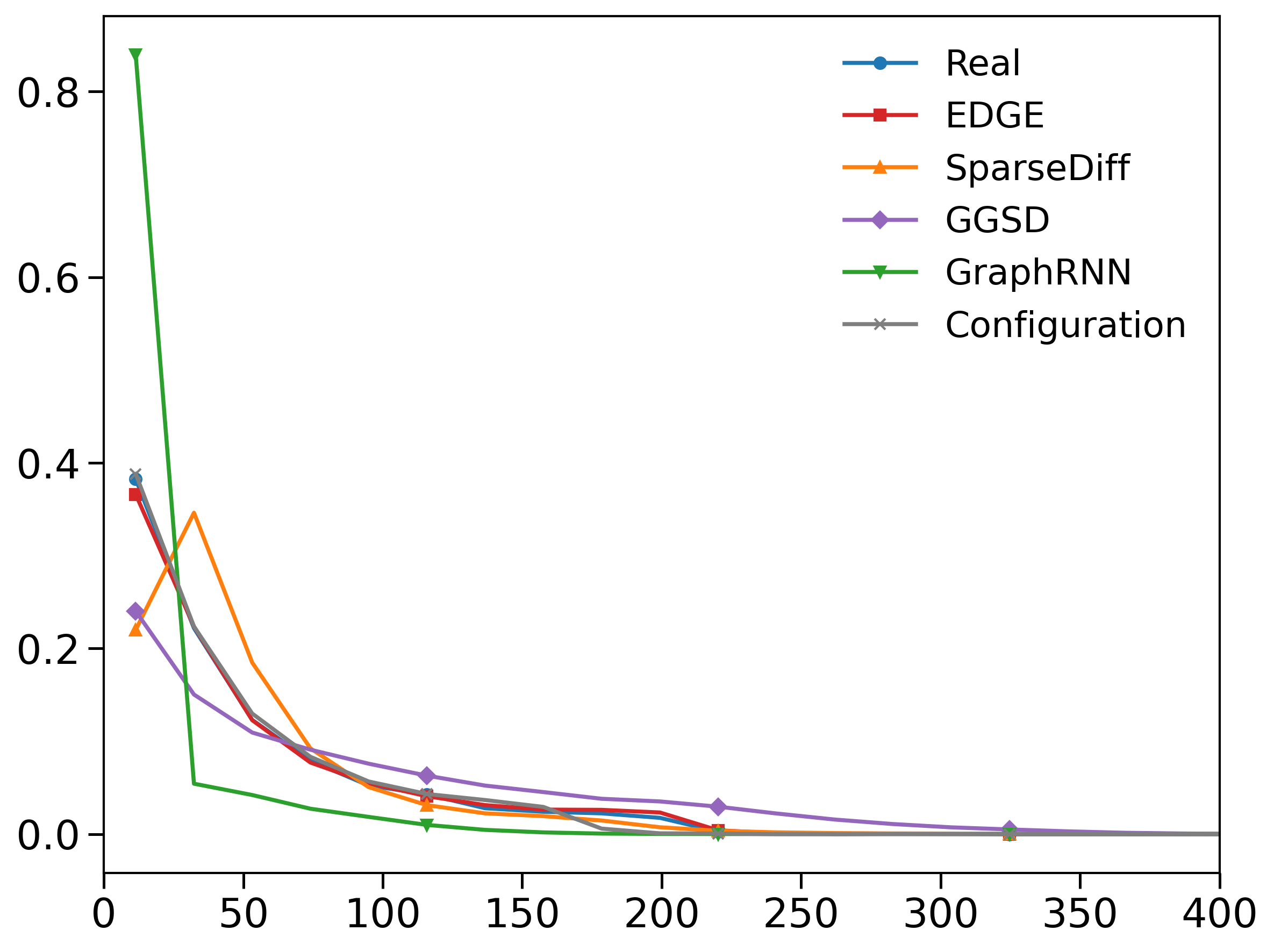}}
    \hfill
    \subfloat[Clustering coefficient]{%
        \includegraphics[width=0.33\textwidth]{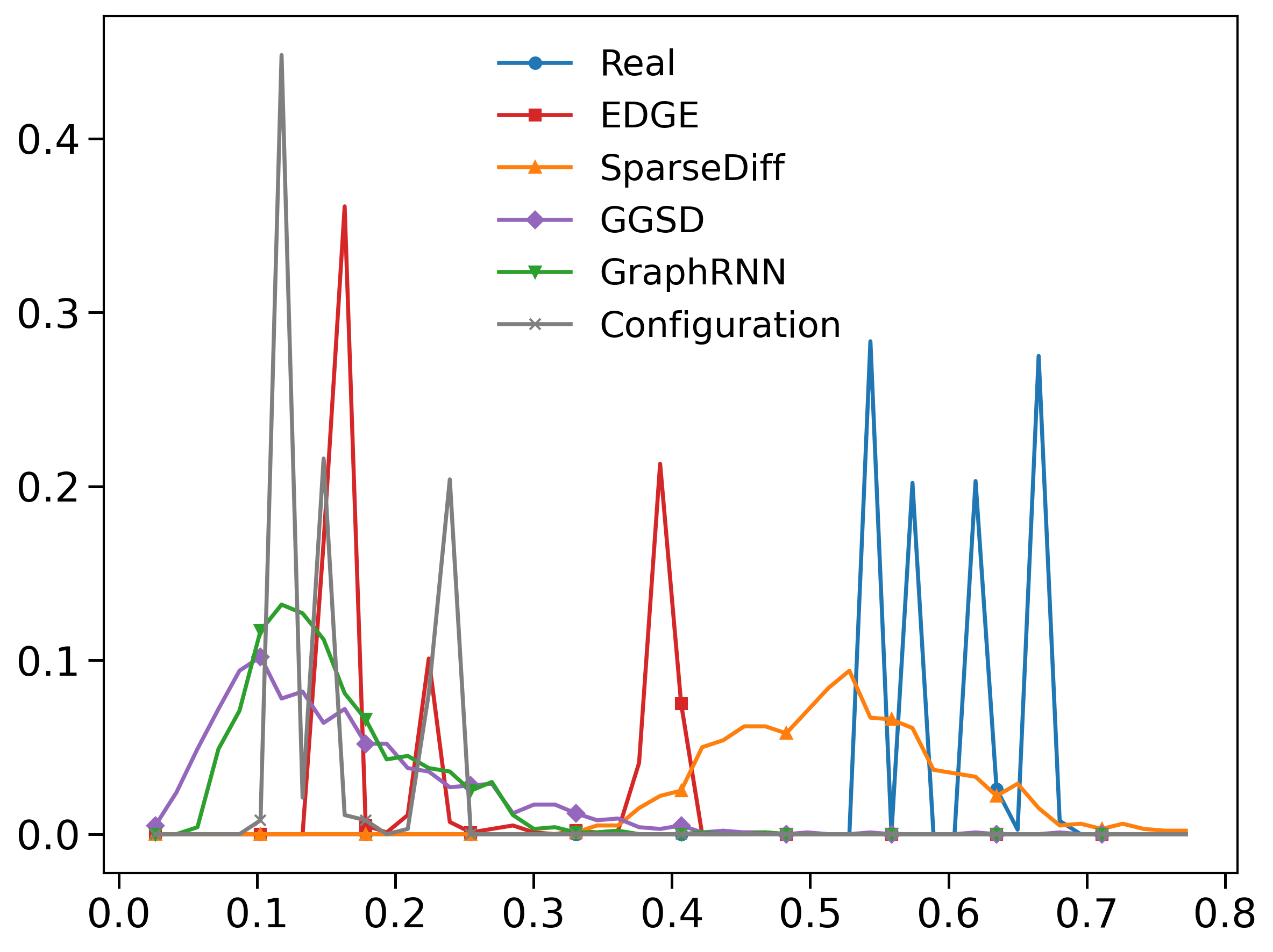}}

    \vspace{6pt}

    \subfloat[Largest eigenvalue]{%
        \includegraphics[width=0.33\textwidth]{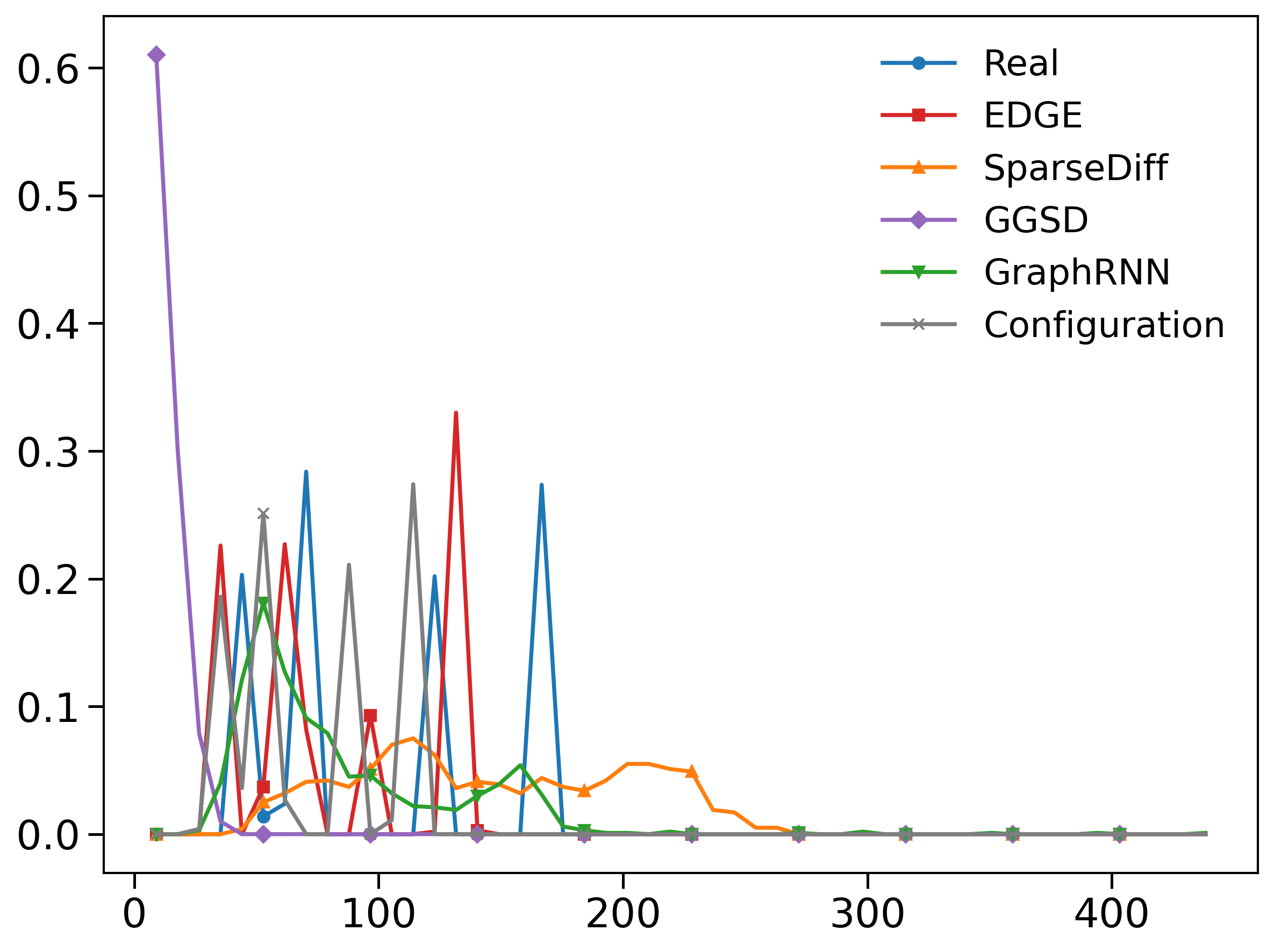}}
    \hfill
    \subfloat[Link density]{%
        \includegraphics[width=0.33\textwidth]{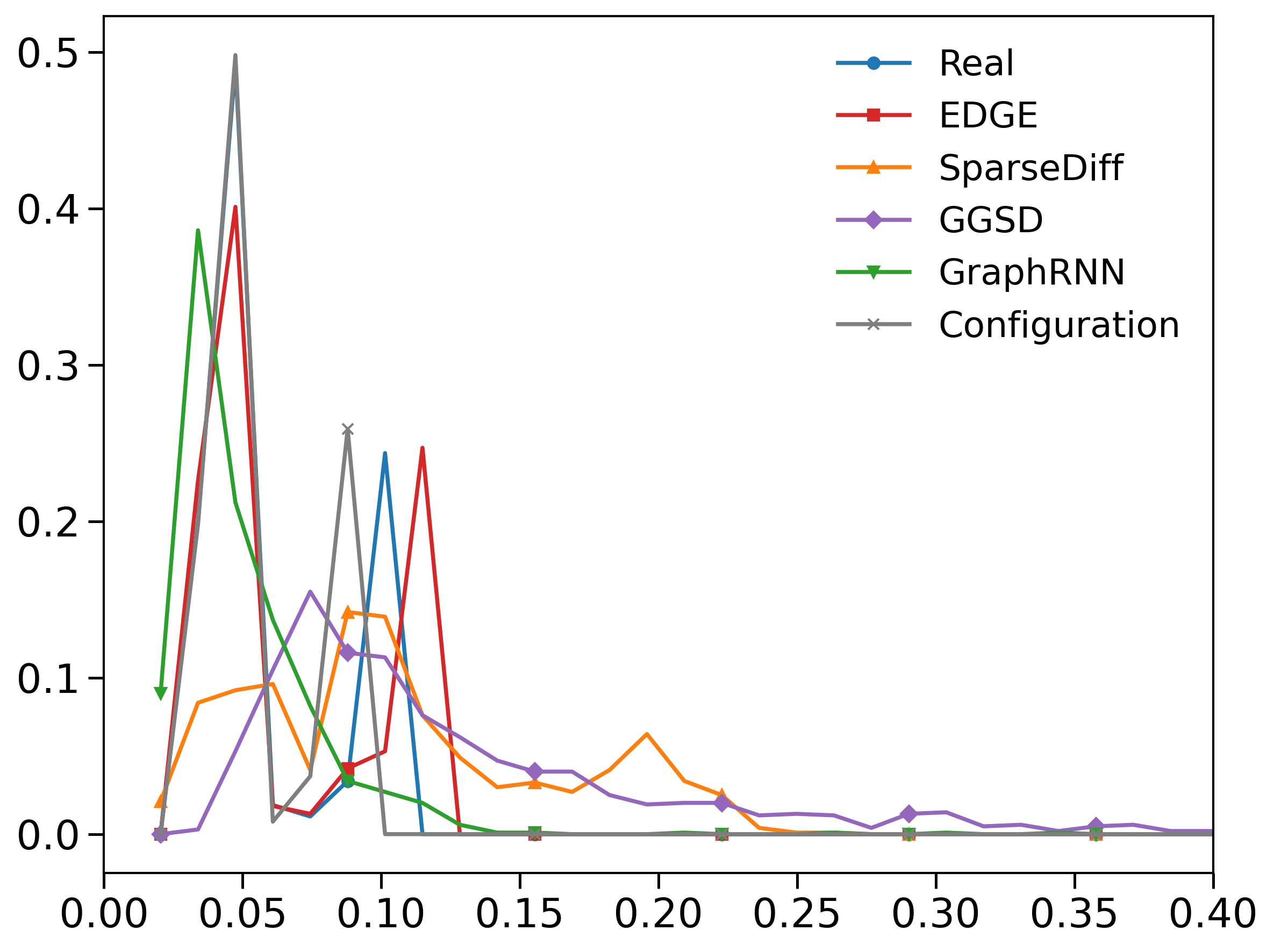}}
        \hfill
    \subfloat[Modularity]{%
        \includegraphics[width=0.33\textwidth]{Subnetworks_statistics/soc-advogato_modularity.png}}
    \caption{Probability distributions of fundamental network properties, including the node number, node degree, clustering coefficient, largest eigenvalue, link density, and modularity for real-world neighborhood networks derived from facebook dataset and the corresponding synthetic networks generated by each model.}
    \label{fig:facebook-distribution}
\end{figure*}

\begin{figure*}[t]
    \centering
    \includegraphics[width=1.0\textwidth]{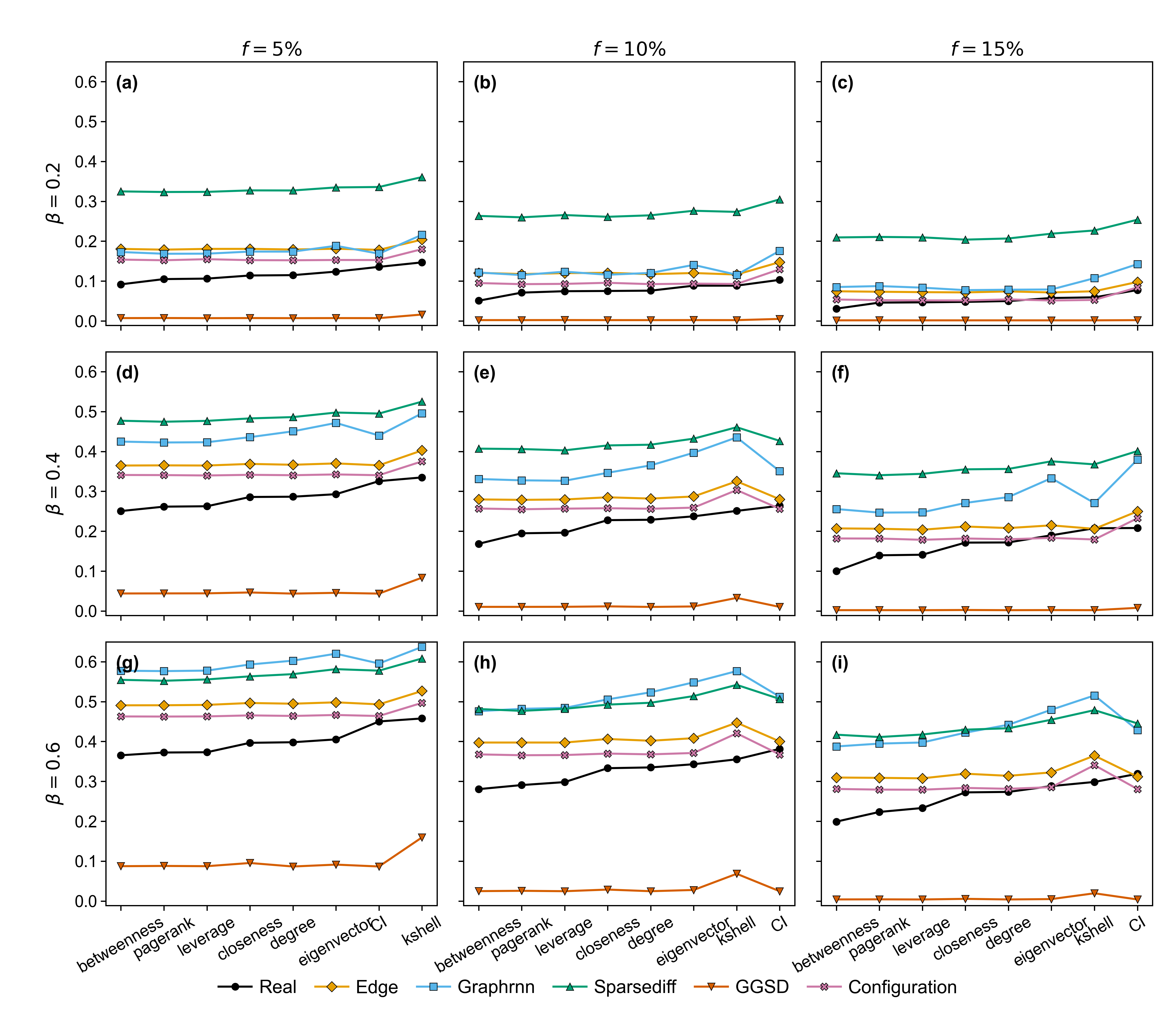}
    \caption{
        Performances of considered immunization strategies across infection rates
        $\beta \in \{0.2,0.4,0.6\}$, evaluated by outbreak size on the
        facebook dataset, for immunization rates
        $f \in \{5\%,10\%,15\%\}$.
    }
    \label{fig:facebook-immunization}
\end{figure*}

\begin{figure*}[t]
    \centering
    \subfloat[Node number]{%
        \includegraphics[width=0.33\textwidth]{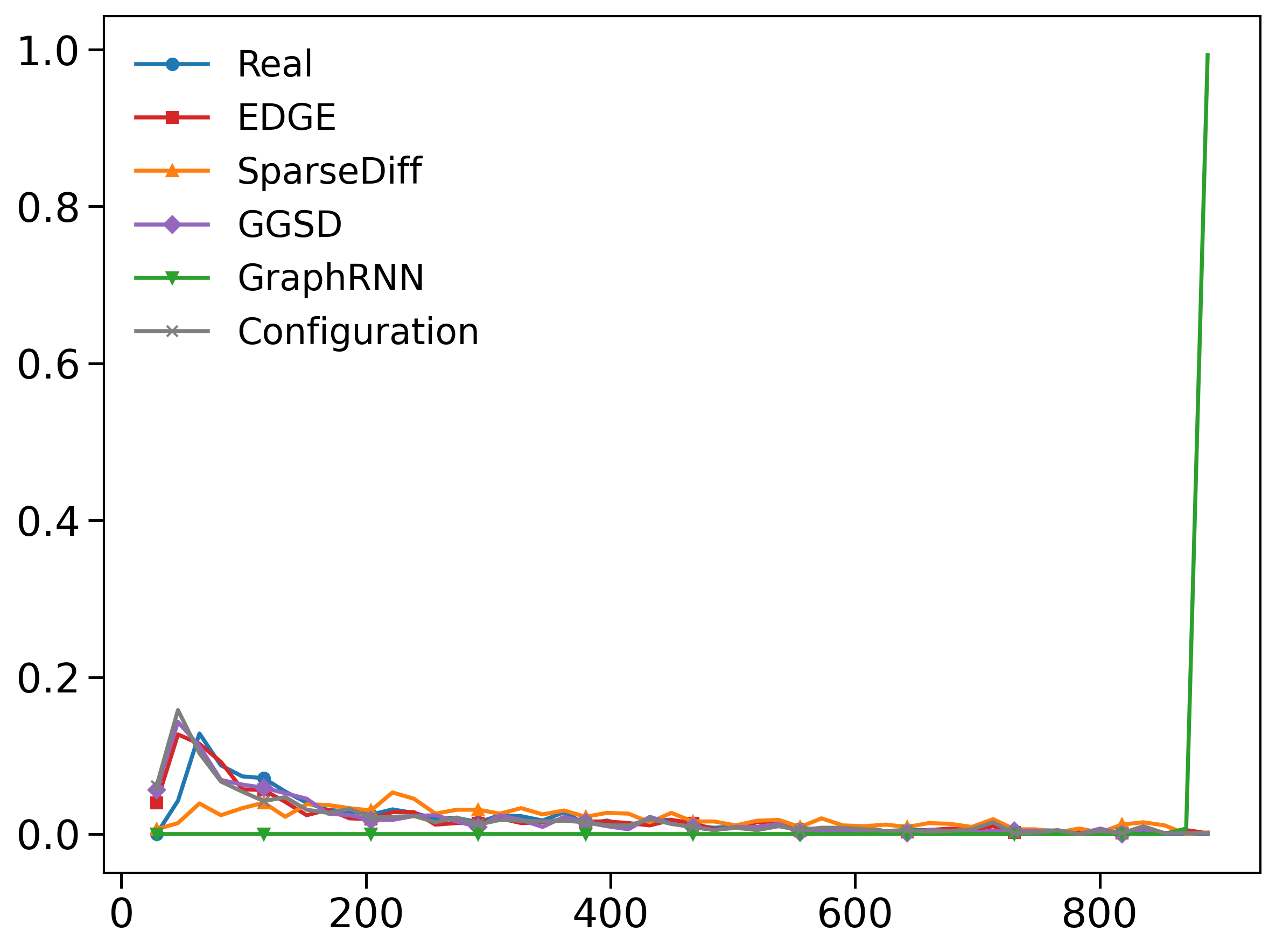}}
    \hfill
    \subfloat[Node degrees]{%
        \includegraphics[width=0.33\textwidth]{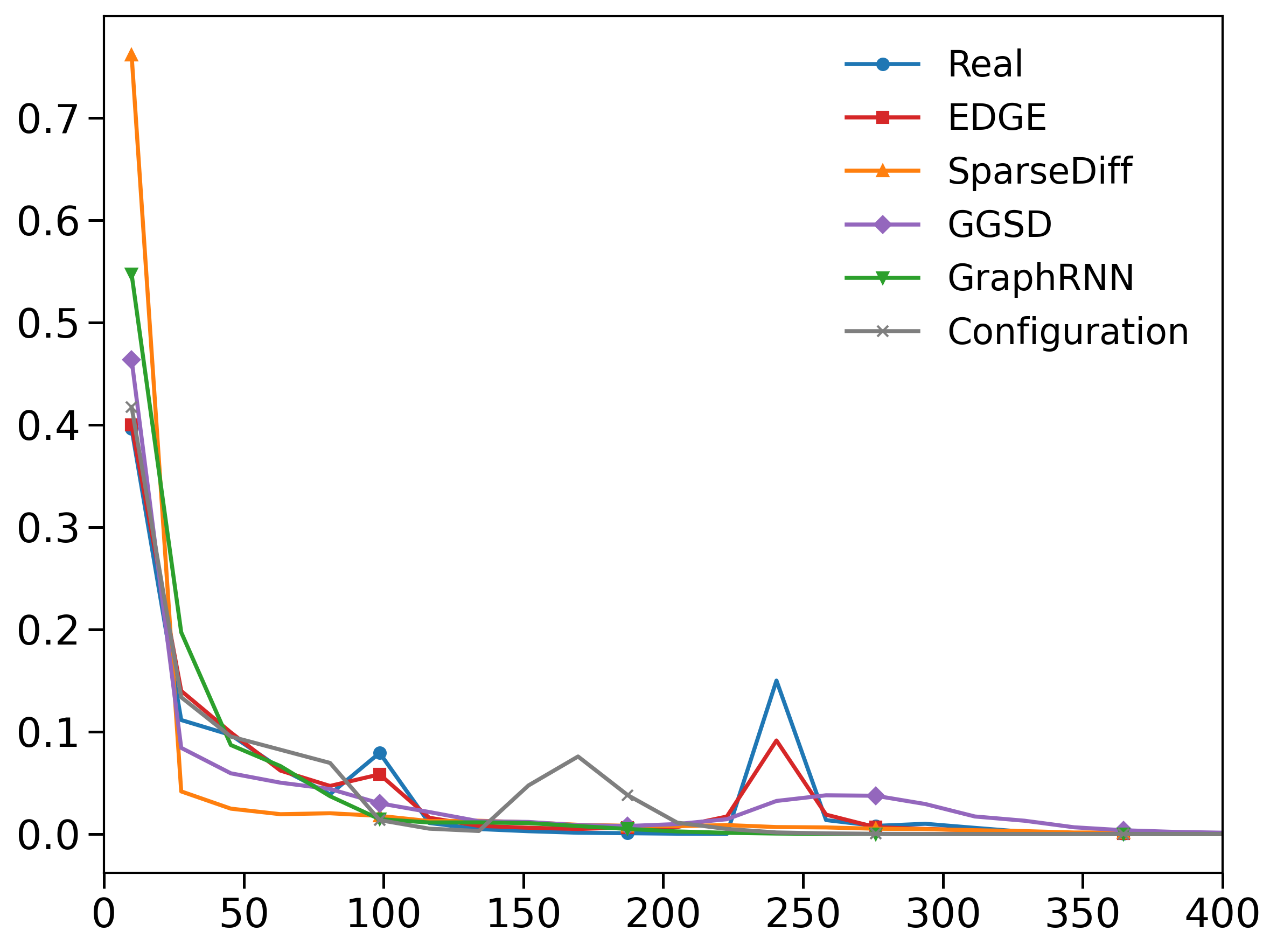}}
    \hfill
    \subfloat[Clustering coefficient]{%
        \includegraphics[width=0.33\textwidth]{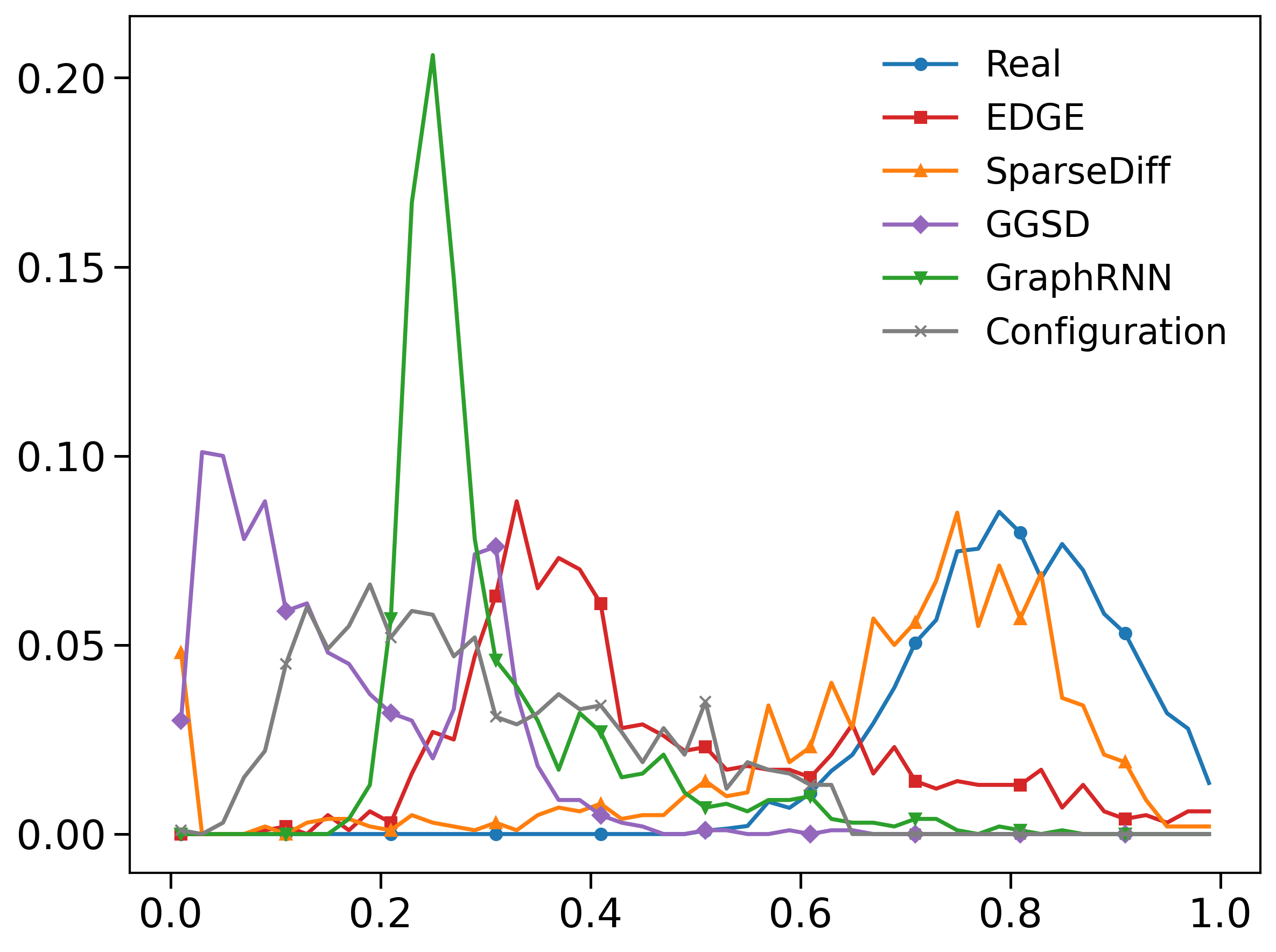}}

    \vspace{6pt}

    \subfloat[Largest eigenvalue]{%
        \includegraphics[width=0.33\textwidth]{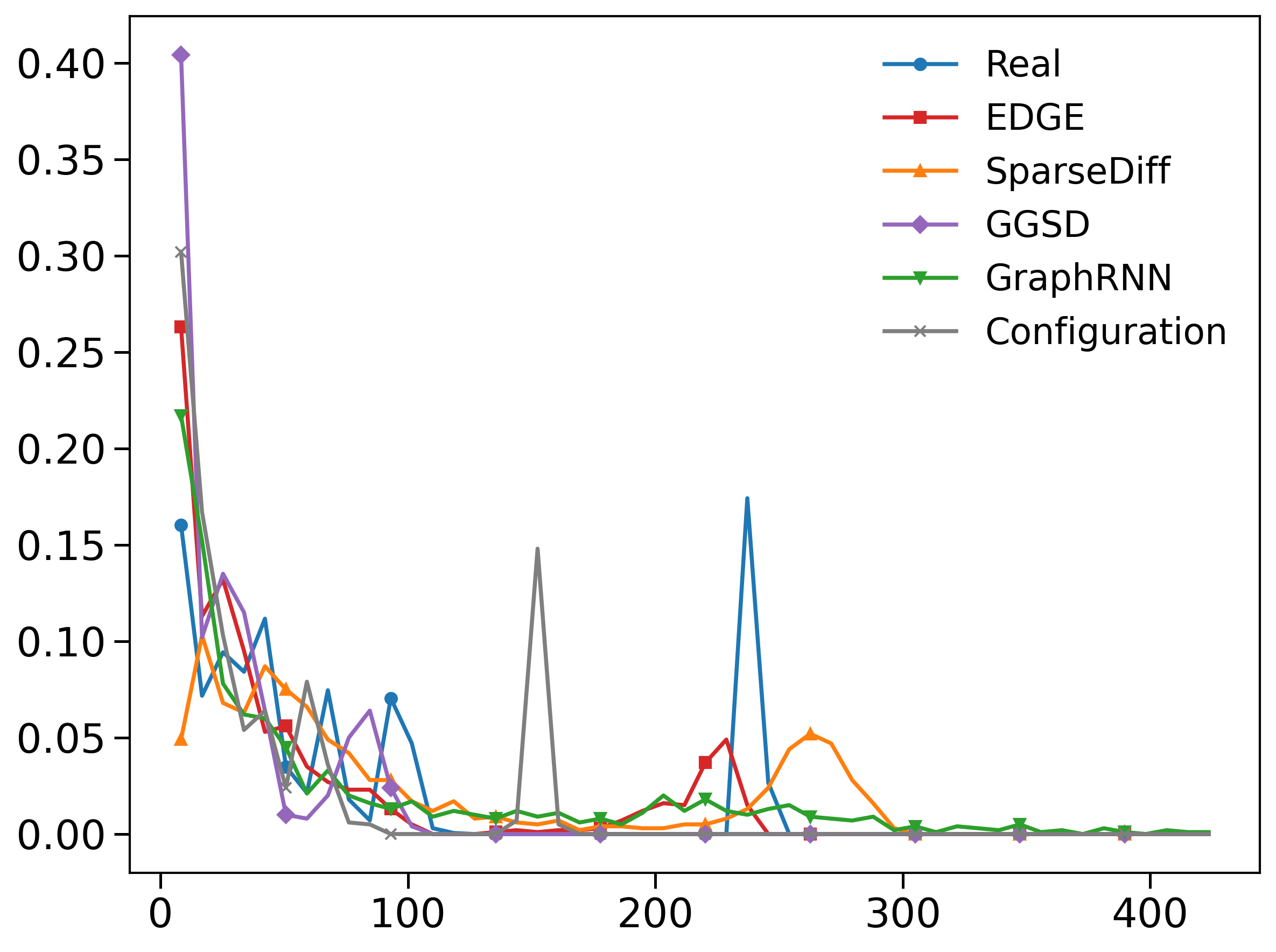}}
    \hfill
    \subfloat[Link density]{%
        \includegraphics[width=0.33\textwidth]{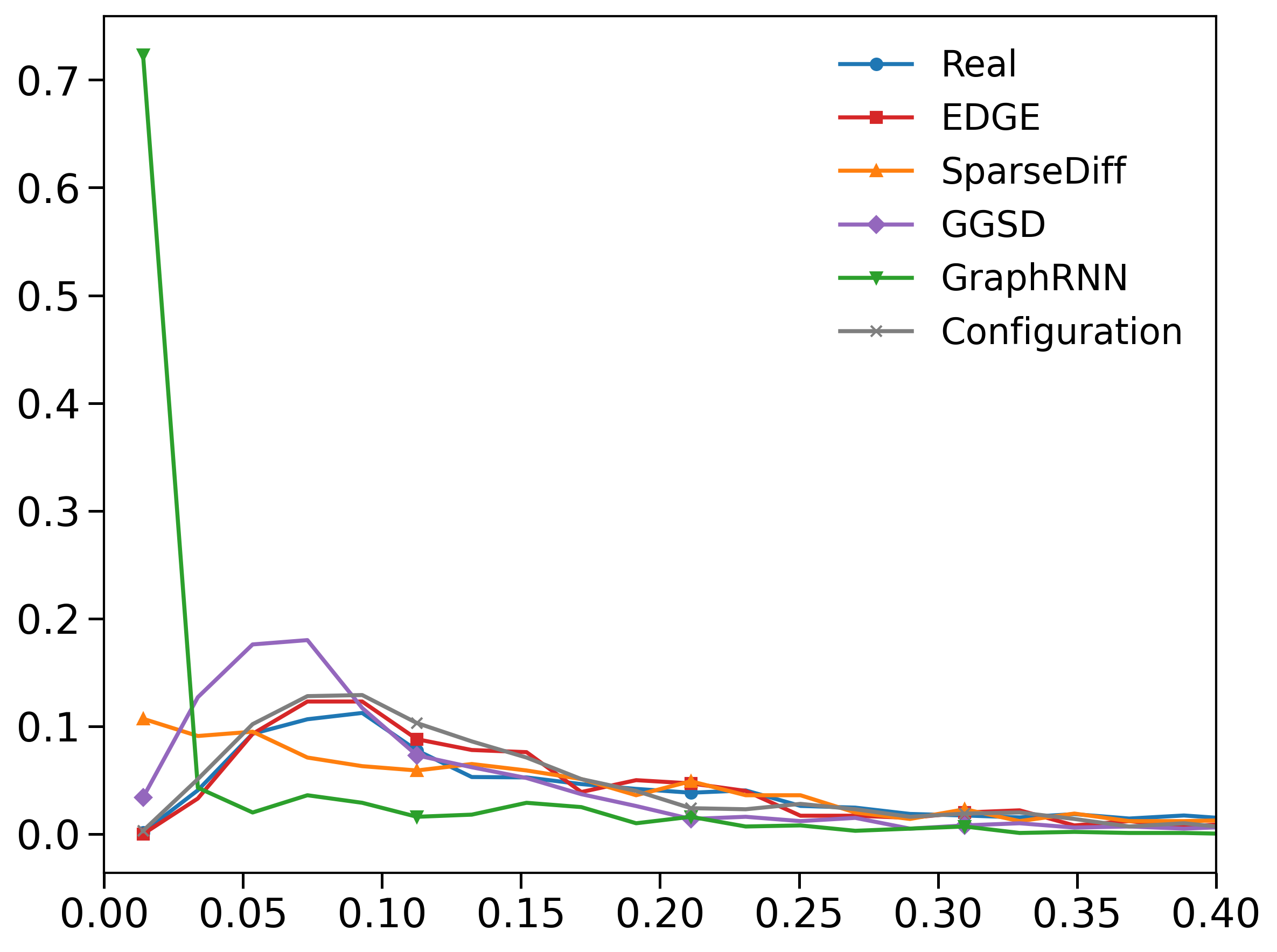}}
        \hfill
    \subfloat[Modularity]{%
        \includegraphics[width=0.33\textwidth]{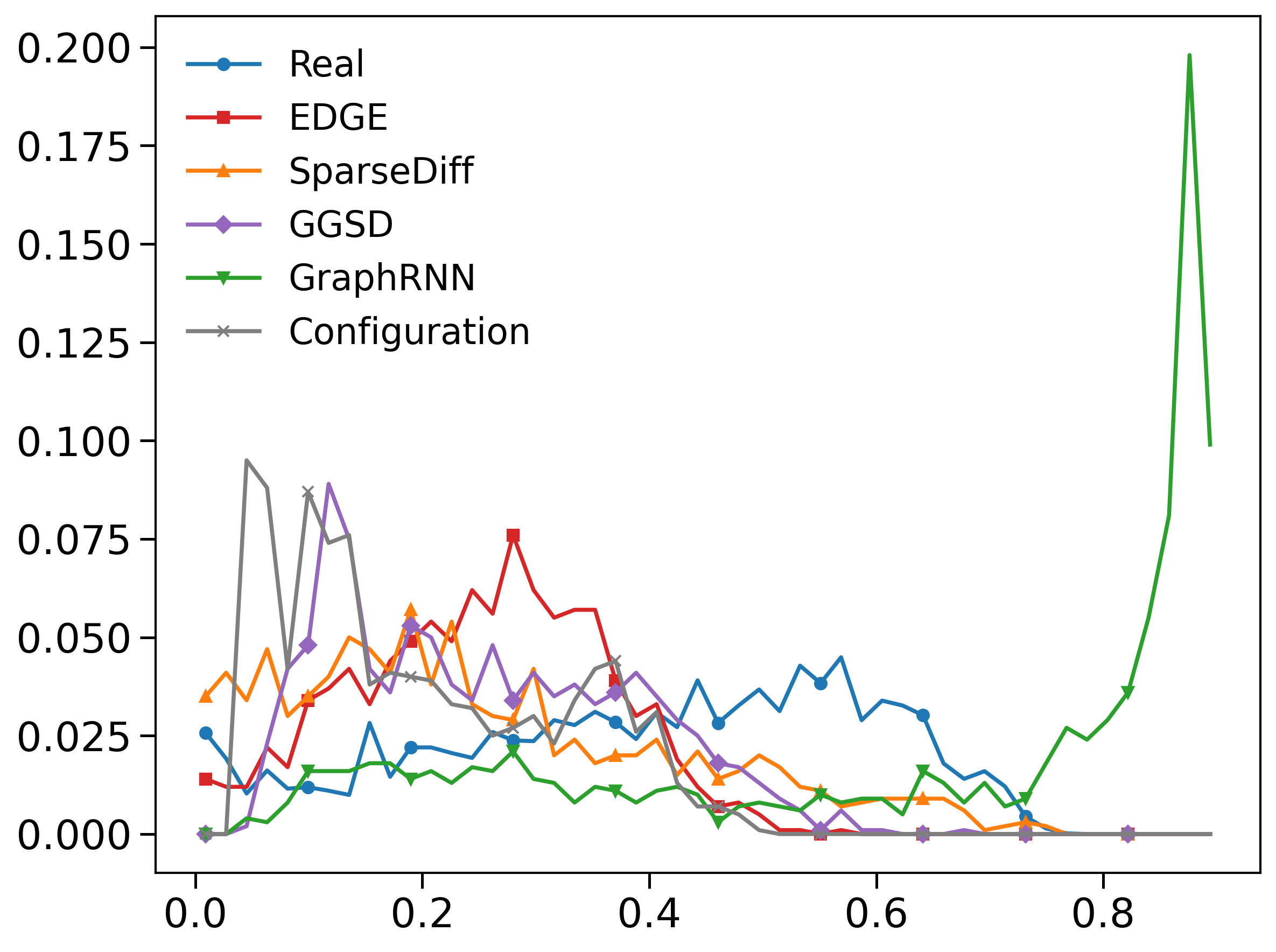}}
    \caption{Probability distributions of fundamental network properties, including the node number, node degree, clustering coefficient, largest eigenvalue, link density, and modularity for real-world neighborhood networks derived from CA-HepPh dataset and the corresponding synthetic networks generated by each model.}
    \label{fig:CA-HepPh-distribution}
\end{figure*}

\begin{figure*}[t]
    \centering
    \includegraphics[width=1.0\textwidth]{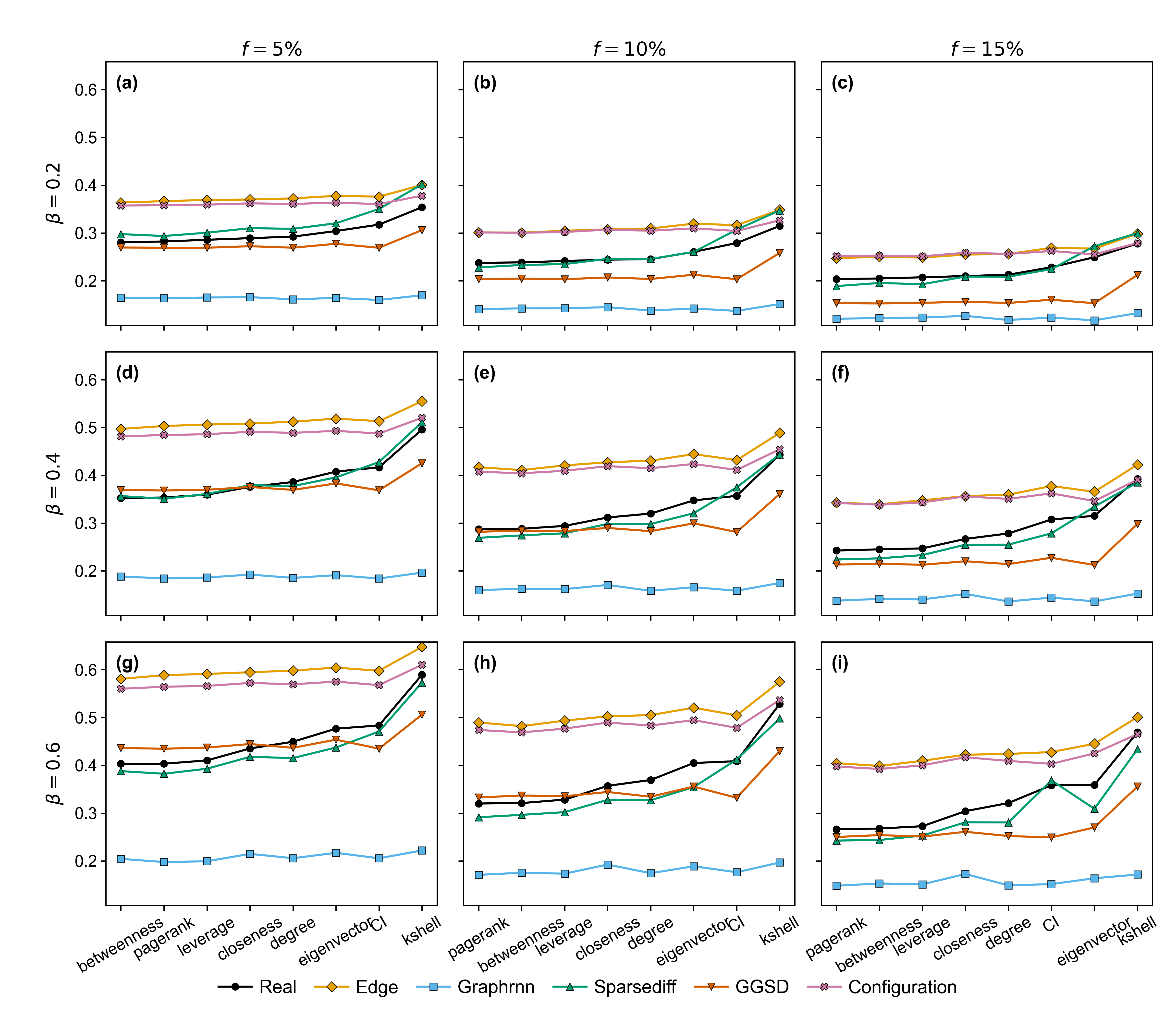}
    \caption{
        Performances of considered immunization strategies across infection rates
        $\beta \in \{0.2,0.4,0.6\}$, evaluated by outbreak size on the
        CA-HepPh dataset, for immunization rates
        $f \in \{5\%,10\%,15\%\}$.
    }
    \label{fig:CA-HepPh-immunization}
\end{figure*}
\end{document}